\documentclass[sigconf]{acmart}

\usepackage{booktabs} 

\usepackage[english]{babel}
\usepackage{moresize}
\usepackage{amsmath}
\usepackage{algorithmic}
\usepackage{balance}
\usepackage{comment}
\usepackage{paralist}
\usepackage{bm}
\usepackage{pgfplots}
\usetikzlibrary{pgfplots.dateplot}

\usepackage{flushend}
\usepackage[english]{babel}
\usepackage[latin1]{inputenc}
\usepackage{mathrsfs}
\usepackage{graphicx}

\usepackage{amssymb}
\usepackage{amsfonts}
\usepackage{url}
\usepackage{longtable}
\usepackage{rotating}
\usepackage{multirow}
\usepackage{mathrsfs}
\usepackage{subfigure}
\usepackage{enumitem}
\usepackage[linesnumbered,algoruled,boxed,lined]{algorithm2e}
\usepackage{adjustbox}
\usepackage{hyperref}
\usepackage{pgfplots}
\usetikzlibrary{pgfplots.dateplot}
\definecolor{tblue}{RGB}{31,119,180}
\definecolor{torange}{RGB}{255,127,14}
\definecolor{tgreen}{RGB}{44,160,44}
\definecolor{tred}{RGB}{214,39,40}
\definecolor{tpurple}{RGB}{148,103,189}

\newcommand{\hide}[1]{} 

\newcommand{\ie}{\emph{i.e.,}\xspace}
\newcommand{\eg}{\emph{e.g.,}\xspace}

\newcommand{\trans}{{\mkern-1.5mu\mathsf{T}}}

\newcommand{\model}{\text{DCRec}}
\newcommand{\paratitle}[1]{\noindent\textbf{#1}}

\pgfplotsset{compat=1.17}

\copyrightyear{2023}
\acmYear{2023}
\setcopyright{acmlicensed}\acmConference[WWW'23]{Proceedings of the ACM Web Conference 2023}{April 30-May 4, 2023}{Austin, TX, USA}
\acmBooktitle{Proceedings of the ACM Web Conference 2023 (WWW'23), April 30-May 4, 2023, Austin, TX, USA}
\acmPrice{15.00}
\acmDOI{10.1145/3543507.3583361}
\acmISBN{978-1-4503-9416-1/23/04}

\begin{document}



\title{Debiased Contrastive Learning for Sequential Recommendation}

\author{Yuhao Yang}
\affiliation{The University of Hong Kong}
\email{yuhao-yang@outlook.com}

\author{Chao Huang}
\authornote{Chao Huang is the corresponding author.}
\affiliation{The University of Hong Kong}
\email{chaohuang75@gmail.com}

\author{Lianghao Xia}
\affiliation{The University of Hong Kong}
\email{aka\_xia@foxmail.com}

\author{Chunzhen Huang}
\affiliation{Wechat, Tencent}
\email{chunzhuang@tencent.com}

\author{Da Luo}
\affiliation{Wechat, Tencent}
\email{lodaluo@tencent.com}

\author{Kangyi Lin}
\affiliation{Wechat, Tencent}
\email{plancklin@tencent.com}
\renewcommand{\shortauthors}{Yuhao Yang et al.}
\begin{abstract}
Current sequential recommender systems are proposed to tackle the dynamic user preference learning with various neural techniques, such as Transformer and Graph Neural Networks (GNNs). However, inference from the highly sparse user behavior data may hinder the representation ability of sequential pattern encoding. To address the label shortage issue, contrastive learning (CL) methods are proposed recently to perform data augmentation in two fashions: (i) randomly corrupting the sequence data (\eg stochastic masking, reordering); (ii) aligning representations across pre-defined contrastive views. Although effective, we argue that current CL-based methods have limitations in addressing popularity bias and disentangling of user conformity and real interest. In this paper, we propose a new \underline{D}ebiased \underline{C}ontrastive learning paradigm for \underline{Rec}ommendation (\model) that unifies sequential pattern encoding with global collaborative relation modeling through adaptive conformity-aware augmentation. This solution is designed to tackle the popularity bias issue in recommendation systems. Our debiased contrastive learning framework effectively captures both the patterns of item transitions within sequences and the dependencies between users across sequences. Our experiments on various real-world datasets have demonstrated that \model\ significantly outperforms state-of-the-art baselines, indicating its efficacy for recommendation. To facilitate reproducibility of our results, we make our implementation of \model\ publicly available at: \url{https://github.com/HKUDS/DCRec}.

\end{abstract}


\maketitle

\section{Introduction}
\label{sec:intro}

Recommender systems (RSs) are increasingly popular in addressing the information overload problem on the Web, especially on platforms such as shopping sites, video platforms, and social networks. These systems help users discover items of interest and enhance their user experience. Among the different types of RSs, sequential recommenders are commonly used to predict future item interactions based on historical behavior sequences~\cite{fang2020deep}.

In the past few years, numerous neural network-based methods have been proposed by researchers to effectively model user interest transitions on item sequences. Examples of such methods include using RNNs~\cite{gru4rec, gru4rec2} or attention-based models~\cite{sasrec, bert4rec} to capture users' evolving interests over time, as reflected by their historical item sequences. However, these methods heavily rely on sufficient interaction data and semantically-rich sequences, making them inadequate for addressing issues such as sparsity~\cite{li2020time}, short sequences~\cite{insert}, and noise~\cite{zhang2021causerec} in recommendation.



As self-supervised learning (SSL) has proven to be effective in the field of recommender systems~\cite{sgl, kgcl, mhcn, lightgcl}, researchers have sought to leverage this paradigm by introducing contrastive learning tasks into sequential recommendation models~\cite{cl4srec, s3rec, duorec, iclrec}. To incorporate supplementary SSL signals,~\cite{cl4srec,s3rec} utilize various data augmentations on sequences or item features to enforce agreement between the augmented views for embedding contrasting. Other methods~\cite{iclrec, duorec} apply contrastive learning by identifying semantically positive pairs for sequences or items for recommendation.

\begin{figure}[t]
\centering
\includegraphics[width=\linewidth]{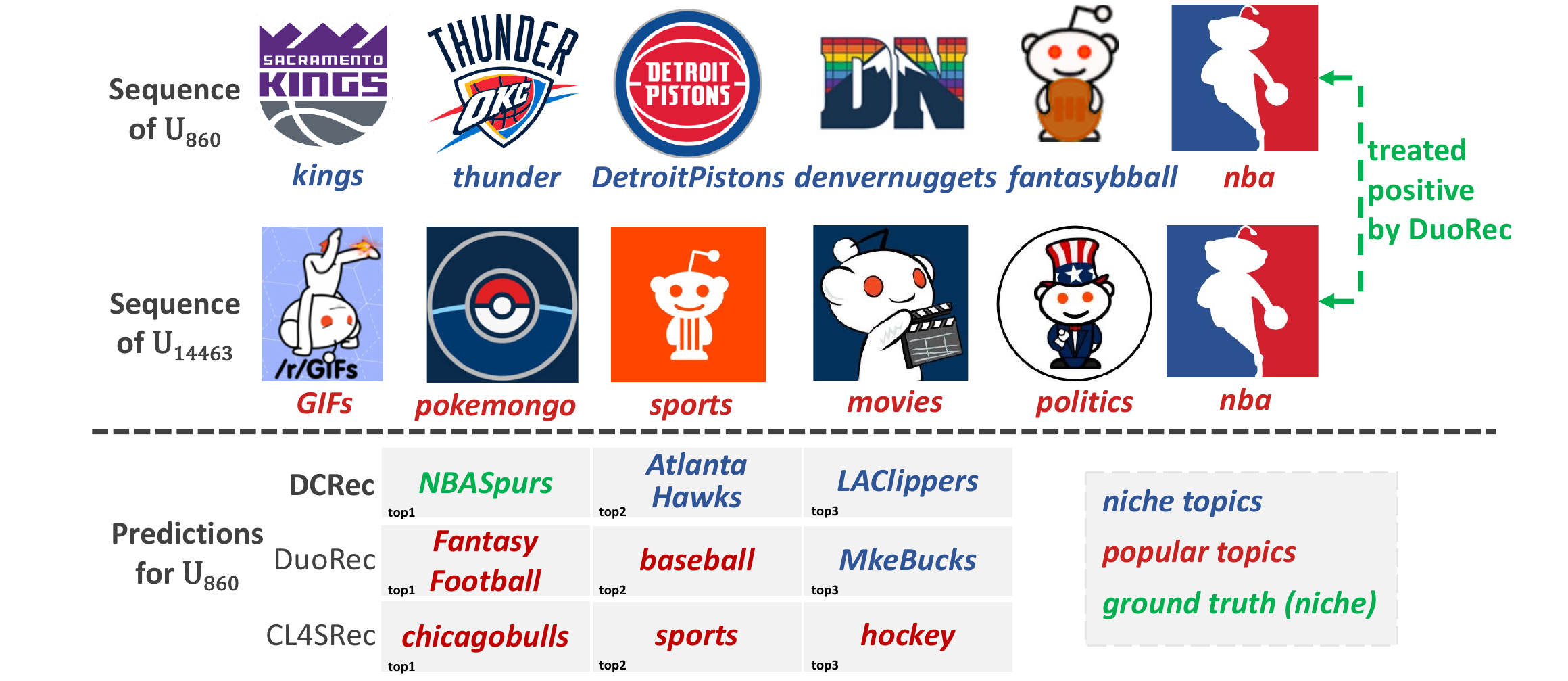}
\vspace{-0.25in}
\caption{A motivating case from the Reddit data illustrates how the lack of attention to popularity bias and user conformity can lead to suboptimal recommendation performance.}
\label{fig:intro_case}
\vspace{-0.22in}
\end{figure}

Although these methods have shown significant improvements in recommendation performance, we believe that existing methods across various research lines have not adequately addressed the inherent popularity bias in data augmentation. To illustrate this issue, we present a case study using data from the Reddit dataset in Figure~\ref{fig:intro_case}. The figure depicts a user, $U_{860}$, who subscribes to a series of niche basketball topics (shown in blue) and also subscribes to the popular topic ``nba'' (shown in red). Another user, $U_{14463}$, subscribes to a range of popular topics observed from his interaction behaviors, with the ``nba'' topic also included. The first user subscribes to ``nba'' due to their genuine interest in basketball sports, while the second user has shown a preference for popular topics in general. For instance, In DuoRec, these two user sequences are wrongly viewed as positive pairs, regardless of the dominance of the user' behaviors, \eg interest or conformity. This, in turn, leads to inaccurate data augmentation for misleading user preference learning.

Moreover, we note that the predictions for $U_{860}$ generated by our proposed \model\ and the state-of-the-art self-supervised sequential models are dissimilar. Our \model\ proves to be effective in capturing the user's interests, enabling accurate ranking of the ground-truth as the top-ranked result. However, the other two models (\ie, DuoRec, CL4SRec) are susceptible to the influence of popularity bias and produce inaccurate recommendations. Therefore, effectively capturing both the interest and conformity components of user intent and modeling them in a disentangled manner, is crucial to enhancing the performance of sequential recommendation against the prevalent popularity bias in data augmentation.

This work proposes \model, a Debiased Contrastive learning framework for sequential Recommendation, to address the limitations mentioned above. Specifically, \model\ integrates contrastive learning with conformity and interest disentanglement to learn augmented representations that are aware of popularity bias. This new paradigm distills informative self-supervision signals for effective augmentation. By integrating relation learning from both sequential and collaborative views, our contrastive learning is conducted across view-specific representations, which can reflect both intra-sequence transitional patterns and inter-sequence global user dependency. Our proposed \model\ disentangles user conformity from noisy item interactions using a multi-channel conformity weighting network, which is based on three semantic channels. The aim of this new approach is to address the issue of bias in the contrastive learning paradigm in recommender systems.


In summary, our work makes the following contributions:

\begin{itemize}[leftmargin=*]

\item We highlight the significance of addressing the popularity bias problem in sparse and noisy user sequence data by extracting self-supervision signals in an adaptable way that disentangles user conformity and actual interest for recommendation.

\item We propose a novel recommendation model, called \model, which addresses the issue of popularity bias in user sequence data through a multi-channel conformity weighting network. Furthermore, our model adapts the strength of contrastive regularization to effectively augment the training data.

\item We demonstrate the effectiveness of our proposed method on several real-world datasets, where our method consistently outperforms state-of-the-art sequential recommendation methods while mitigating the effects of popularity bias.

\end{itemize}

    

    

\vspace{-0.1in}
\section{Methodology}
\label{sec:solution}

\begin{figure*}
    \centering
    \includegraphics[width=\linewidth]{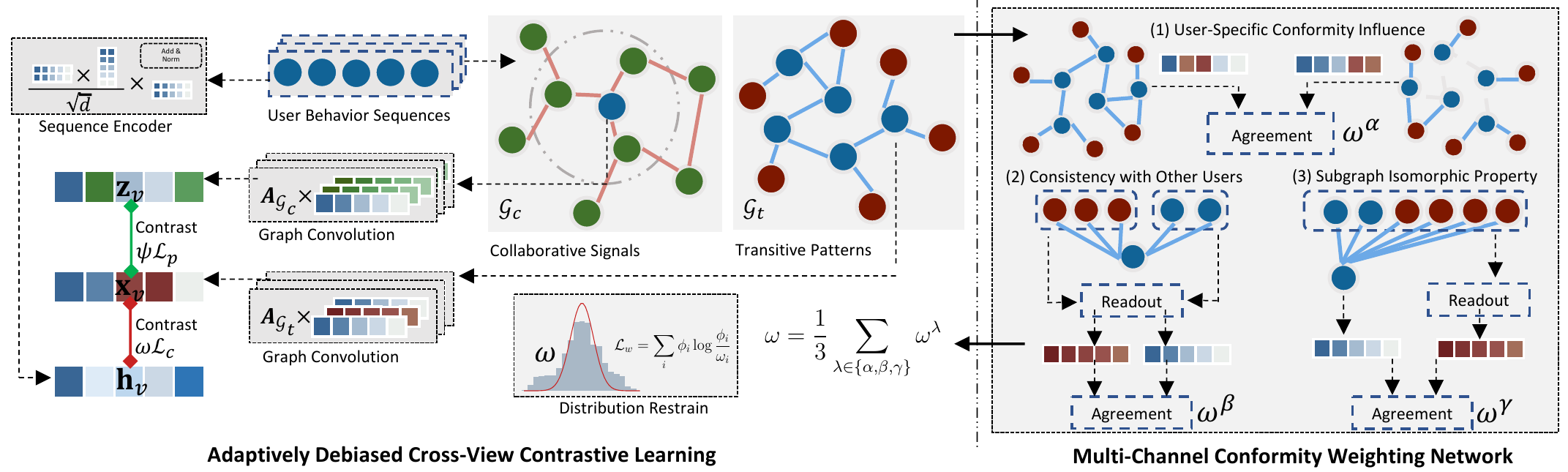}
    \vspace{-0.2in}
    \caption{The overall framework of \model. $\mathcal{G}_c$ and $\mathcal{G}_t$ are built to encode the sequences from diversified views (left part). In addition, we generate reasonable interaction-level conformity weights $\omega$ from the rich structure of $\mathcal{G}_t$ (right part). The weights are restrained in normal distribution and empower the cross-view contrastive learning to be adaptive and aware of conformity.}
    \label{fig:arch}
    \vspace{-0.15in}
\end{figure*}

The overall model architecture of our \model\ is shown in Figure~\ref{fig:arch}.

\subsection{Task Formulation}
\noindent \textbf{Notations}. We suppose a recommender with a set of users and items denoted by $\mathcal{U} (u\in \mathcal{U})$ and $\mathcal{V} (v\in \mathcal{V})$, respectively. For each user, his/her engaged subset of items in a temporal order is defined as $\boldsymbol{s}_u=\left(v_1, v_2, \cdots, v_T\right)$. Here, $T$ is the sequence length which varies by users, and indexed by $t$, \ie $1\leq t \leq T$. Following settings in~\cite{bert4rec,iclrec}, we conduct the padding operation over different item sequences ($\boldsymbol{s}_u\ \in \mathcal{S}$) to mitigate the variable length.\\\vspace{-0.12in}

\noindent \textbf{Task}. 
Our objective is to develop a personalized ranking function that takes into account the past item sequences of a user, and predicts the next item ($v_{T+1}$) that the user is most likely to adopt.


\vspace{-0.1in}
\subsection{Sequential Pattern Encoding}
As of now, Transformer has emerged as the dominant method for encoding sequences, capable of mapping temporally-ordered tokens from different types of sequential data to latent representation space. Examples include textual data~\cite{devlin2018bert} and electronic health data~\cite{poulain2021transformer}. Our sequential pattern encoder is built upon the Transformer, inspired by the effectiveness of this approach in modeling item sequence in~\cite{bert4rec,wu2020sse,yuan2022multi}. This allows us to incorporate temporal context into embeddings, resulting in an effective representation of the user's sequential behavior.


We start by adding a positional embedding $\mathbf{p}_v$ to the initial item representation $\mathbf{v}_v$ using the operation $\mathbf{h}_v^0 = \mathbf{v}_v \oplus \mathbf{p}_v$, which serves as the input item embedding $\mathbf{h}_v^0$ for the first block of Transformer. We represent each user's item sequence with an embedding matrix $\mathbf{H}_u^0 \in \mathbb{R}^{T \times d}$, where $T$ is the length of the sequence and $d$ is the dimension of the item embedding. The embedding matrix corresponds to the padded item sequence $\boldsymbol{s}_u$ of the user. To capture the correlations between items, we apply a self-attention layer with multi-head ($N$) channels to the user's item embedding matrix:
\begin{align}
    \text{MH}\left({\textbf{H}_u^\ell}\right) &= \left(\text{head}_1 \mathbin\Vert \text{head}_2 \mathbin\Vert  \cdots \mathbin\Vert \text{head}_N\right)\mathbf{W}^D \\
	\text{head}_n &= \text{Attention}\left( \textbf{H}_u^\ell \mathbf{W}^Q_n, \textbf{H}_u^\ell \mathbf{W}^K_n,  \textbf{H}_u^\ell \mathbf{W}^V_n \right),
\end{align}
\noindent $\mathbf{W}^Q_n, \mathbf{W}^K_n, \mathbf{W}^V_n \in \mathbb{R}^{d \times d/N}$ represents the head-specific mapping matrices corresponding to the query, key, value dimension, respectively. $\mathbf{W}^D \in \mathbb{R}^{d \times d}$ is a learnable projection matrix, and $\textbf{H}_u^\ell$ is the embedding matrix of user $u$'s sequence $\boldsymbol{s}_u$ at the $\ell$-th block of Transformer. Here, the self-attention calculation is conducted as: $\text{Attention}\left( \mathbf{Q},\mathbf{K},\mathbf{V}  \right) = \text{softmax}\left( \frac{\mathbf{Q}\cdot \mathbf{K}^\trans}{\sqrt{d/N}} \right)\mathbf{V}$. $\frac{d}{N}$ is the scale factor.


To inject non-linearity into the embedding generation, a point-wise feed-forward network (FFN) is used for representation transformation within the sequential pattern encoder, which is defined:
\begin{align}
    \label{eq:transformer}
    \text{PFFN}\left(\mathbf{H}_u^{\ell}\right) &= [\text{FFN}\left(\mathbf{h}_1^{\ell}\right)^\trans, \cdots, \text{FFN}\left(\mathbf{h}_T^{\ell}\right)^\trans] \\
	\text{FFN}\left(\mathbf{x}\right) &= \text{GELU}\left(\mathbf{x}\mathbf{W}_1^{\ell} + \mathbf{b}_1^{\ell}\right)\mathbf{W}_2^{\ell}+\mathbf{b}_2^{\ell},
\end{align}
\noindent where $\mathbf{W}_1^{\ell}, \mathbf{W}_2^{\ell}, \mathbf{b}_1^{\ell}, \mathbf{b}_2^{\ell}$ are learnable model parameters as projection and bias terms. $\text{GELU}(\cdot)$ is the activation function.

\subsection{Unifying Sequential and CF Views}
In real-life applications, long-tail sequences with a limited number of items are prevalent in recommendation scenarios~\cite{liu2020long,cl4srec}. These sequences pose challenges to most existing solutions. In particular, short sequences with very few items can hardly provide sufficient contextual signals for neural sequence encoders. This issue affects various types of models, such as self-attention mechanisms~\cite{bert4rec, sasrec}, and graph neural networks~\cite{gcsan, srgnn, mbht, surge}. To tackle the challenge of short sequences with very few items in sequential recommenders, we propose to unify the sequential view of item transitions and the collaborative view of user-item interactions. This design aims to capture the implicit cross-sequence user dependencies, allowing user-wise knowledge transfer in sequential recommender systems. This aspect is largely overlooked in most current solutions.


To achieve the goal of unifying the sequential view of item transitions and the collaborative view of user-item interactions, you can start by generating two graphs: \emph{item transition graph} $\mathcal{G}_t$ and \emph{item co-interaction graph} $\mathcal{G}_c$. To be specific, $\mathcal{G}_t$ and $\mathcal{G}_c$ over the item set $\mathcal{V}$ are constructed by following the instructions below:
\begin{itemize}[leftmargin=*]
\item \textbf{Item Transition Graph $\mathcal{G}_t$}. To capture the transitional relationships among items from the sequential pattern view, adjacent item pairs (\eg $v_{t-1}$, $v_{t}$) in each sequence $\boldsymbol{s}_u$ are connected with an edge in $\mathcal{G}_t$. Given the item sequences of all users $\mathcal{S} = \{ \boldsymbol{s}_1, \boldsymbol{s}_2, \cdots, \boldsymbol{s}_{|U|} \}$, the adjacency matrix $\mathbf{A}_{\mathcal{G}_t} \in \mathbb{R}^{|\mathcal{V}| \times |\mathcal{V}|}$ representing the item correlations in graph $\mathcal{G}_t$ is generated by:
\begin{align}
\label{eq:gt}
    \mathbf{A}^u_{\mathcal{G}_t}(v_p, v_q) = \begin{cases}
                        1, & |p-q|=1 \\
                        0, & \text{otherwise}
    \end{cases} ;\quad
    \mathbf{A}_{\mathcal{G}_t} = \sum_{u=1}^{|U|}\mathbf{A}^u_{\mathcal{G}_t},
\end{align}
\noindent where $\mathbf{A}^u_{\mathcal{G}_t}$ denotes the user-specific item transition connections over sequence $\boldsymbol{s}_u$. Here, $p$ and $q$ denotes the position index in sequence. We sum up $\mathbf{A}^u_{\mathcal{G}_t}$ of all users ($u\in \mathcal{U}$) to obtain $\mathbf{A}_{\mathcal{G}_t}$. The adjacency matrix $\mathbf{A}_{\mathcal{G}t}$ takes into account the transition frequency between items with edge weights in the item transition graph. \\\vspace{-0.12in}


\item \textbf{Item Co-Interaction Graph $\mathcal{G}_c$}. To incorporate collaborative signals to model the cross-user dependencies, we generate another graph $\mathcal{G}_c$ to maintain the item correlations based on their co-interaction patterns. To this end, we firstly construct the interaction matrix $\mathbf{R} \in \mathbb{R}^{|\mathbf{U}| \times |\mathbf{V}|}$ between users and items by setting the entry $\mathbf{R}_{u, v}=1$ if user $u$ has adopted item $v$ and $\mathbf{R}_{u, v}=0$ otherwise. With the operation $\mathbf{A}_{\mathcal{G}_c} = \mathbf{R}^\trans\mathbf{R}$, we obtain the initial correlation strength between items in $\mathbf{A}_{\mathcal{G}_c}$ based on their co-interaction frequency. To filter out less-relevant item-wise connections, we apply \emph{top}-$k(\cdot)$ function to keep highly-relevant connections among items in $\mathbf{A}_{\mathcal{G}_c}$ based on top-$k$ co-interaction frequency of each item. Here, $k$ determines the density of $\mathbf{A}_{\mathcal{G}_c}$.


\end{itemize}

After generating the item transition graph $\mathcal{G}_t$ and co-interaction graph $\mathcal{G}_c$, we utilize the graph neural network to project individual item into latent embedding space. Formally, our graph convolution-based message passing is presented as follows:
\begin{equation}
    \label{eq:gcn}
    \mathbf{X}^{(l+1)} = \left(\mathbf{D}_t^{-\frac{1}{2}} \mathbf{A}_{\mathcal{G}_t} \mathbf{D}_t^{-\frac{1}{2}}\right)\mathbf{X}^{(l)};\ 
    \mathbf{Z}^{(l+1)} = \left(\mathbf{D}_c^{-\frac{1}{2}} \mathbf{A}_{\mathcal{G}_c} \mathbf{D}_c^{-\frac{1}{2}}\right)\mathbf{Z}^{(l)}
\end{equation}
\noindent We let $\mathbf{X}^{(l)}$ and $\mathbf{Z}^{(l)}$ respectively denote the embedding matrix of items over the item transition graph ($\mathcal{G}_t$) and the co-interaction graph ($\mathcal{G}_c$) under the $l$-th graph layer. $\mathbf{D}_a$ and $\mathbf{D}_i$ are degree matrices used for graph normalizing. To simplify the model with lightweight GNN architecture, we remove the redundant transformation and activation operations during the message propagation.

\subsection{Adaptive Cross-View Contrastive Learning}
\label{sec:adaptive}
Building on the success of contrastive data augmentation across various domains, including vision learning~\cite{he2020momentum}, text mining~\cite{rethmeier2021primer}, and graph modeling~\cite{zhu2021graph}, our \model\ method harnesses self-supervised signals through contrastive learning across different item semantic views. Nonetheless, the popularity bias is often overlooked, as conformity can entangle real interests and subsequently influence user behaviors~\cite{zheng2021disentangling,chen2021autodebias}. For instance, a user might be influenced by conformity to click on a product or watch a short video, following the actions of others, rather than being genuinely interested in the content. If user interest and conformity are not disentangled when generating augmented signals, contrastive learning methods may focus on incorrect positive pairs, thereby introducing biased information. This can lead to less-interested recommendation.


Intuitively, conformity may vary across users and interactions. For example, user conformity and real interest might be entangled in a complex manner, jointly driving interaction behaviors. This complexity makes it challenging to accurately disentangle conformity from genuine interest, which is essential for providing more helpful augmented SSL signals. To address this challenge, we propose a debiased cross-view contrastive learning approach with adaptive augmentation that incorporates interaction-level conformity. We develop a multi-channel conformity weighting network (CWNet) to calculate the conformity degree of an interaction. By incorporating the estimated conformity degrees into our contrastive learning paradigm, we can adaptively determine the regularization strength. This allows the model to more effectively disentangle user interests from conformity behaviors.


\subsubsection{\bf Multi-Channel Conformity Weighting Network}
In our CWNet module, we aim to learn the conformity degree of an interaction between user $u$ and item $v$ from three semantic channels.\\\vspace{-0.12in}

\begin{itemize}[leftmargin=*]
\item (1) \textbf{ User-Specific Conformity Influence}. First, we propose to infer the interaction-level (\eg $u-v$) conformity degree by considering the conformity of user $u$ based on his/her past interactions. Given a user with strong conformity, their interactions are more likely to be influenced by popularity bias compared to others who exhibit strong individuality. To obtain the conformity degree of user $u$, we perturb the item transition graph $\mathcal{G}_t$ by removing the edges generated from $u$'s sequence $\boldsymbol{s}u$. This results in the generation of an augmented adjacency matrix $\bar{\mathbf{A}}{\mathcal{G}c}$, where $\bar{\mathbf{A}}^u{\mathcal{G}_t}(v_p, v_q) = 0$ for any two adjacent items $v_p$ and $v_q$ in $\boldsymbol{s}_u$. Subsequently, both the original and augmented item transition graphs are fed into our graph encoder (as per Eq.~\ref{eq:gcn}) to generate two embeddings ($\mathbf{x}_v, \mathbf{x}_v^\prime$) for the target item $v$. The user-specific conformity influence, denoted as $\omega^1_{\left(u,v\right)}$, is estimated using the cosine similarity between the two embeddings ($\mathbf{x}_v, \mathbf{x}v^\prime$), calculated as $\omega^{\alpha}{\left(u,v\right)} = \cos\left( \mathbf{x}_v, \mathbf{x}_v^\prime \right)$. A larger $\omega^u$ score indicates that user $u$'s interactions have little influence over the item graph structures, suggesting that their interaction patterns are more likely to be observed from others, \ie strong user conformity.  \\\vspace{-0.12in} 

\item (2) \textbf{Consistency with Other Users}. We also propose to calculate the conformity from the perspective of considering the sequential behavior consistency between the target user and others. In particular, for a given $u-v$ interaction, we compare the learned transitional patterns of user $u$ with those of other relevant users. To be specific, given the target item $v$, we aggregate the intra-sequence neighboring information using mean-pooling among inner neighbors within the sequence $\boldsymbol{s}_u$. The overall transitional patterns of other correlated users are combined to obtain $\overline{\mathbf{x}}_{O_v}$, which is derived from $v$'s outer neighbors $O_v$ across different user sequences. After that, the transition consistency is measured by $\omega^{\beta}_{\left(u,v\right)} = \cos\left(\overline{\mathbf{x}}_{N_v}, \overline{\mathbf{x}}_{O_v}\right)$. This measure quantifies the degree of consistency between the target user's sequential behavior and that of other users, providing insights into conformity. \\\vspace{-0.12in}

\item (3) \textbf{Subgraph Isomorphic Property}. The isomorphic property of item subgraph is also an important factor in reflecting user conformity with similar interaction patterns. To incorporate this factor into our conformity estimation, we calculate the similarity between item $v$'s embedding $\mathbf{x}_v$ and the representation $\overline{\mathbf{x}}_{O_v}$ aggregated from its outer neighbors, \ie $\omega^{\gamma}_{\left(u,v\right)} = \cos\left(\mathbf{x}_v, \overline{\mathbf{x}}_{O_v}\right)$.
\end{itemize}

\noindent \textbf{Mixing Signals from Different Channels.} We derive the final interaction-level conformity degree by fusing the information from the above three channels. Here, we first adopt mean-pooling over channel-specific results as: $\omega_{\left(u,v\right)} = \frac{1}{3}\sum_{\lambda \in \left\{\alpha,\beta,\gamma \right\}}\omega_{\left(u,v\right)}^{\lambda}$. Following the mapping strategy in~\cite{kgcl,zhu2021graph}, we perform the transformation for $\omega$ values as follows:
\begin{equation}
\label{eq:omega}
    \omega^{(1)} = \text{sigmoid}\left(\omega\right);\ 
    \omega^{(2)} = \frac{\omega^{(1)} - \omega_{min}^{(1)}}{\omega_{max}^{(1)} - \omega_{min}^{(1)}};\ 
    \omega^{(3)} = \frac{\mu_c}{\overline{\omega}^{(2)}} \cdot \omega^{(2)}
\end{equation}
\noindent $\mu_c$ is the hyperparameter that adjusts the mean value $\overline{\omega}$ of $\omega$. We omit the subscript $\left(u,v\right)$ for simplicity and adopt $\omega = \omega^{(3)}$ as the output conformity. Furthermore, to approximate the conformity degrees with normal distribution, we adopt the KL-divergence over the derived conformity results of all interactions:
\begin{equation}
    \label{eq:kl}
    \mathcal{L}_w = \sum_{i=1}^{|\{(u,v)\}|} \phi_i\log\frac{\phi_i}{\omega_i},
\end{equation}
\noindent where $\phi_i$ is generated by random sampling from normal distribution with the hyperparameter $\mu_c$ for the mean and $\sigma$ for the standard deviation. $\omega_i$ is the conformity weighting result.

\subsubsection{\bf Conformity-aware Contrastive Augmentation}
To enhance our \model\ with adaptively debiased augmentation, we integrate the conformity factor into our embedding contrasting paradigm to determine the agreement regularization strength. As discussed before, both sequential and collaborative views are generated through different encoders, \ie Transformer and GNNs. Our \model\ employs contrastive learning (CL) to learn conformity-aware augmented representations from two key dimensions:\\\vspace{-0.12in}

\noindent \textbf{Contrasting from User Dimension}. The first stage of our CL paradigm aims to realize the knowledge transfer across different users. By contrasting user-specific preferences with cross-user global interaction patterns, the learned augmented representations can naturally preserve user-wise implicit dependencies. In this process, the conformity regularizer weakens the impacts of perturbations caused by popularity bias for SSL augmentation. Given the embedding $\mathbf{h}_v$ and $\mathbf{x}_v$ encoded generated by our sequential pattern encoder (Eq.~\ref{eq:transformer}) and transition graph encoder (Eq.~\ref{eq:gcn}), respectively, our debiased contrastive learning paradigm is given as follows:
\begin{equation}
\label{eq:cl1}
    \mathcal{L}_u = -\sum_{u\in \mathcal{U}}\sum_{v \in \boldsymbol{s}_u}  \omega_{\left(u,v\right)} \log \frac{\exp\left(\cos\left(\mathbf{h}_v, \mathbf{x}_v\right)/\tau\right)}{\sum_{v^\prime\in \mathcal{V}} \exp\left(\cos\left(\mathbf{h}_v, \mathbf{x}_{v^\prime}\right)/\tau\right)},
\end{equation}
\noindent In the SSL loss $\mathcal{L}_u$, InfoNCE~\cite{infonce} is adopted for embedding contrasting. By incorporating our learned conformity factor $\omega$, we allow representations $\mathbf{h}_v$ and $\mathbf{x}_v$ to supervise each other adaptively, that is, weighted by the interaction-level conformity. \\\vspace{-0.12in}


\noindent \textbf{Contrasting from Item Dimension}. The goal of our second stage CL is to extract self-supervision signals by contrasting the global item embedding $\mathbf{x}_v$ with the item semantic representation $\mathbf{z}_v$. Our conformity factor $\omega_{u,v}$ is incorporated into this contrasting process by estimating the uniformity $\psi_{\left(u,v\right)}=1-\omega_{\left(u,v\right)}$. Formally, our item dimension CL loss $\mathcal{L}_v$ is defined as follows:
\begin{equation}
\label{eq:cl2}
    \mathcal{L}_v = -\sum_{u\in \mathcal{U}}\sum_{v \in \boldsymbol{s}_u}  \psi_{\left(u,v\right)} \log \frac{\exp\left(\cos\left(\mathbf{x}_v, \mathbf{z}_v\right)/\tau\right)}{\sum_{v^\prime\in \mathcal{V}} \exp\left(\cos\left(\mathbf{x}_v, \mathbf{z}_{v^\prime}\right)/\tau\right)},
\end{equation}
\noindent In our CL paradigm, instance self-discrimination~\cite{sgl,xia2022hypergraph} is used for generating positive pairs. Representation of different samples are pushed apart as negative pairs to reflect embedding uniformity.

\subsection{Model Training and Prediction}
In the training phase, the last interacted item of each sequence $\boldsymbol{s}_u$ is regarded as the label for model optimization. In the prediction phase, to encourage the cooperation between sequence and collaborative views, we combine view-specific item embeddings into an aggregated representation $\mathbf{p}_v$ with the learnable attentive weights:
\begin{align}
\label{eq:fusion}
    f\left(\mathbf{e}\right) = \frac{\exp\left(\mathbf{a}^\trans \cdot \mathbf{W}_a \mathbf{e}\right)}{\sum_{i=1}^3 \exp\left(\mathbf{a}^\trans \cdot \mathbf{W}_a \mathbf{e}\right)};~~~ \mathbf{p}_v = \sum_{\mathbf{e}\in \left\{\mathbf{h},\mathbf{x},\mathbf{z} \right\}}f\left(\mathbf{e}\right)\mathbf{e}
\end{align}
\noindent where $\mathbf{a}\in\mathbb{R}^{d}$ and $\mathbf{W}_a\in\mathbb{R}^{d\times d}$ are trainable attention parameters. Input embedding $\mathbf{e}$ is selected from the set of view-specific representations, \ie $\mathbf{e} \in \left\{\mathbf{h}_v, \mathbf{x}_v, \mathbf{z}_v \right\}$. $\mathbf{p}_v$ is derived through attentive aggregation with the view-specific importance score $f\left(\mathbf{e}\right)$.

The next item interaction probability $\hat{y}_{u,v}$ is derived as $\hat{y}_{u,v} = \mathbf{p}_{|\boldsymbol{s}_u|}^\trans\mathbf{v}$, where we adopt hidden states of the last item on the sequence as the user embedding. For each user and the ground truth item $v_t$ pair, we utilize the cross-entropy as the loss:
\begin{equation}
    \mathcal{L}_{rec} = \sum_{(u,v_{T+1})\in \mathcal{D}^+} -\log\frac{\exp \hat{y}_{u,v_{T+1}}}{\sum_{v^\prime \in \mathcal{V}}\exp\hat{y}_{u,v^\prime}},
\end{equation}
\noindent where $\mathcal{D}^+$ is the training data set of positive interactions at the $T+1$ timesteps.
To supplement the recommendation loss $\mathcal{L}_{rec}$ with our augmented SSL tasks under a multi-task training framework, we define our joint optimized objective $\mathcal{L}$ as:
\begin{equation}
\label{eq:all}
    \mathcal{L} = \mathcal{L}_{rec} + \lambda_1\left(\mathcal{L}_u+\mathcal{L}_v\right) + \lambda_2\left(\mathcal{L}_w\right),
\end{equation}
\noindent where $\lambda_1$ and $\lambda_2$ are parameters to balance the tasks-specific loss. $\mathcal{L}_w$ is the regularization term with KL-divergence for mixing signals (Eq.~\ref{eq:kl}) in our multi-channel conformity weighting network.\\\vspace{-0.12in}

\noindent \textbf{Time Complexity Analysis}. In our sequential pattern encoder, the computational cost is $O\left(T^2d+Td^2\right)$ where the majority of the cost  is attributed to the item-wise self-attention operations. In our GNN encoder, the graph convolutional message passing and aggregation have a complexity of $O\left(|\mathcal{V}|d^2\right)$. In the cross-view representation aggregation, our \model\ requires a computational cost of $O(d^2)$ for attentional weighting. Owing to the independent nature of our sequential and collaborative relation encoders, the Transformer and GNN encoding can be performed in parallel using the CUDA infrastructure for speeding up computation. In summary, the time complexity of our \model\ is $O\left(\left(|\mathcal{V}|+1\right)d^2\right)$, making it comparable to the state-of-the-art GNN-based sequential recommenders. 


\vspace{-0.1in}
\subsection{Theoretical Analyzes of \model}
\begin{figure}
    \centering
    \includegraphics[width=0.9\linewidth]{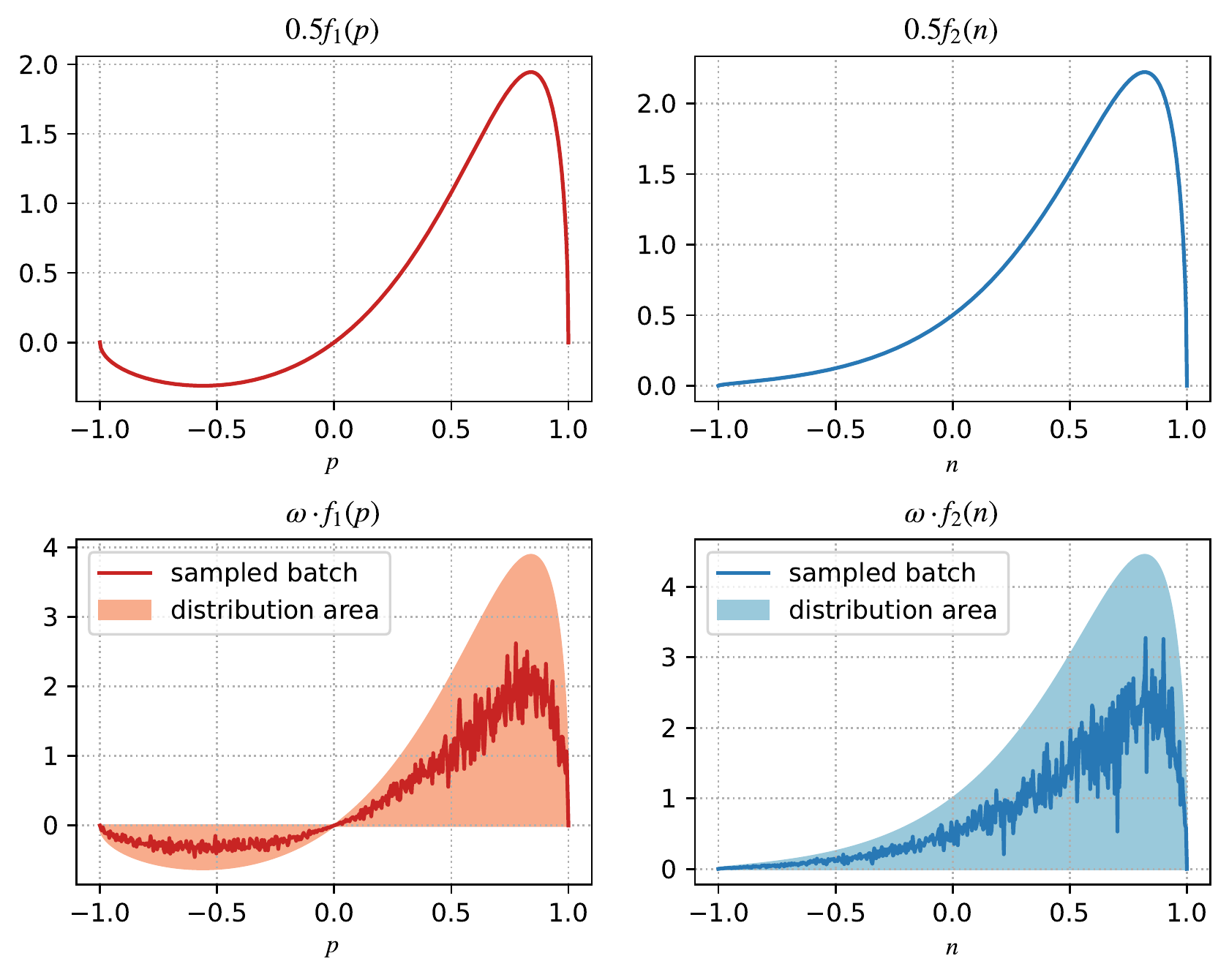}
    \vspace{-0.2in}
    \caption{Upper part: curve of $0.5f(p)$ and $0.5f(n)$ under $\tau=0.4$. Lower part: distribution area of potential values of $\omega \cdot f(p)$ and $\omega \cdot f(n)$ and random samples within a batch.}
    \label{fig:theo}
    \vspace{-0.2in}
\end{figure}

In this section, we provide an analysis of how the new conformity-aware contrastive learning paradigm benefits the recommendation task. We focus on how to bring theoretical interpretability for the conformity-aware adaptive contrastive learning in Equation~\ref{eq:cl1}-\ref{eq:cl2}. We take Equation~\ref{eq:cl1} for studying because of the symmetry of these two equations. Following the discussion in~\cite{kgcl, sgl, khosla2020supervised}, the gradient of the contrastive objective in Equation~\ref{eq:cl1} can be expressed as:
\begin{equation}
\label{eq:contrib1}
    \nabla \mathcal{L}_u^{(u,v)} = \frac{1}{\tau \|\mathbf{h}_v\|}\left(c(v)+\sum_{v^\prime\in V\setminus\{p\}}c(v^\prime)\right),
\end{equation}
\noindent where $\mathcal{L}_u^{(u,v)}$ is the contrastive loss $\mathcal{L}_u$ for an user-item pair $(u,v)$. $c(v)$ and $c(v^\prime)$ are the gradient contribution from the positive pair $(\mathbf{h}_v,\mathbf{x}_v)$ and the negative pair, respectively. Formally, $c(v)$ and $c(v^\prime)$ are derived using the following formulas:
\begin{equation}
\label{eq:contrib2}
    \begin{aligned}
    c(v) &= \left(\mathbf{\bar{x}}_v - \left(\mathbf{\bar{h}}_v^\trans\mathbf{\bar{x}}_v\right)\mathbf{\bar{h}}_v\right)^\trans\left(P_{vv}-1\right)\\
    c(v^\prime) &= \left(\mathbf{\bar{x}}_{v^\prime} - \left(\mathbf{\bar{h}}_v^\trans\mathbf{\bar{x}}_{v^\prime}\right)\mathbf{\bar{h}}_v\right)^\trans P_{vv^\prime},
    \end{aligned}
\end{equation}
\noindent where $P_{vi} = \exp\left(\mathbf{\bar{h}}_v^\trans\mathbf{\bar{x}}_i / \tau\right) \big/ \sum_{i\in V\setminus\{v\}}\exp\left(\mathbf{\bar{h}}_v^\trans\mathbf{\bar{x}}_i / \tau \right)$. $\mathbf{\bar{x}}, \mathbf{\bar{h}}$ are normalized representations. To this end, we can derive two functions $f(p)$ and $f(n)$ that are proportional to the $L_2$ norm of $c(v)$ and $c(v^\prime)$~\cite{sgl}. Specifically, we have the following derivations:
\begin{equation}
\label{eq:contrib3}
    f_1(p) = \sqrt{1-p^2}\left(\exp\left(\frac{p}{\tau}\right)-1\right);\ f_2(n) = \sqrt{1-n^2}\left(\exp\frac{n}{\tau}\right)
\end{equation}
\noindent where $p = \mathbf{h}_v^\trans\mathbf{x}_v$ is the agreement between the positive pair. $n = \mathbf{h}_v^\trans\mathbf{x}_v^\prime$ denotes the similarity between the negatives. To visualize the impact of $\mathcal{L}_u$ without adaptive weight $\omega$, we plot the curve of $0.5f_1(p)$ and $0.5f_2(n)$ in Figure~\ref{fig:theo}. Note that without $\omega$, the coefficient of $\mathcal{L}_u$ is 0.5 by default. From the curves, it is obvious that the contribution of positive and negative samples at different similarity levels are fixed. This means that the model has difficulty in discriminating among diverse samples. In the context of an interest-driven interaction, it is crucial to dynamically reduce the influence of samples from the conformity modeling view.


At this stage, we investigate the advantages of introducing a conformity-aware weight (denoted by $\omega$) in contrastive learning. Specifically, the conformity-aware weight $\omega$ influences the learning process by directly scaling the gradient values. Recall that the distribution of $\omega$ is restrained by normal distribution in Equation~\ref{eq:kl}. The distribution range of $\omega\cdot f_1(p)$ and $\omega\cdot f_2(n)$ creates an area rather than a single curve. We further plot the distribution areas in the lower part of Figure~\ref{fig:theo}. The values are weighted by the interaction-level conformity, falling within the ranges of $(0,f_1(p))$ and $(0,f_2(n))$ following a normal distribution. We also plot the discrete distribution of $\omega\cdot f_1(p)$ and $\omega\cdot f_2(n)$ by sampling two batches of training data. As evident from the distributions, the effect of some samples is enhanced while the influence of others are weakened. This endows the learning process with richer semantics, allowing for a dynamic and adaptive contribution of samples to the contrastive learning gradients with data debiasing. The analyzes also apply to $\mathcal{L}_v$ in Equation \ref{eq:cl2}, since $\gamma = 1-\omega$ has similar properties.


\section{Evaluation}
\label{sec:eval}

\begin{table}[t]
    \centering
    \caption{Detailed statistics of experimental datasets}
    \vspace{-0.15in}
	\resizebox{\linewidth}{!}{
    \begin{tabular}{lcccc}
    \toprule
    Statistics & Reddit & Beauty & Sports & Movielens-20M\\
    \midrule
    \# Users & 14,487 & 56,849 & 85,227 & 96,727 \\
    \# Items & 15,417 & 41,533 & 56,975 & 10,155\\
    \# Interactions & 28,972 & 113,696 & 170,452 & 193,452\\
    \# Avg. Length & 20.95 & 3.67 & 3.70 & 18.20\\
    \# Density & 1e-4 & 5e-5 & 4e-5 & 2e-4\\
    \bottomrule
    \end{tabular}
    }
    \label{tab:dataset}
    \vspace{-0.2in}
\end{table}

\begin{table*}[h]
    \centering
    \caption{Overall performance evaluation across all methods. The best and second best performance are denoted in bold and underline separately. $^\ast$ indicates that the best performance is statistically significant at $p<0.01$ compared to the second best.}
    \vspace{-0.15in}
    \label{tab:results}
    \resizebox{\linewidth}{!}{
    \begin{tabular}{clccccccccccccr}
    \toprule
    Dataset & Metric & Caser & GRU4Rec & SASRec & BERT4Rec & SR-GNN & GCSAN & SURGE & S$^3$-Rec & CL4SRec & DuoRec & ICLRec & \textbf{\model} & \textit{\#Improve}\\
    \midrule
    \multirow{5}{*}{Reddit} & HR@1 & 0.0842 & 0.0858 & 0.0989 & 0.1058 & 0.1741 & 0.1405 & 0.1782 & 0.0187 & 0.1742 & \underline{0.1747} & 0.0532 & \textbf{0.1883}$^\ast$ & 7.78\%\\
    ~ & HR@5 & 0.1410 & 0.1449 & 0.2438 & 0.2566 & \textbf{0.3984} & 0.3037 & 0.3895 & 0.0608 & 0.3762 & 0.3792 & 0.0919 & \underline{0.3899} & -2.18\%\\
    ~ & HR@10 & 0.2121 & 0.2241 & 0.3445 & 0.3820 & \textbf{0.5130} & 0.4154 & 0.4939 & 0.1100 & 0.4925 & 0.4950 & 0.1499 & \underline{0.5057} & -1.44\%\\
    ~ & NDCG@5 & 0.1124 & 0.1145 & 0.1725 & 0.1825 & \underline{0.2903} & 0.2248 & 0.2935 & 0.0396 & 0.2796 & 0.2818 & 0.0725 & \textbf{0.3007}$^\ast$ & 3.58\% \\
    ~ & NDCG@10 & 0.1351 & 0.1398 & 0.2049 & 0.2225 & \underline{0.3273} & 0.2607 & 0.3238 & 0.0533 & 0.3170 & 0.3191 & 0.0909 & \textbf{0.3358}$^\ast$ & 2.60\%\\
    \midrule
    \multirow{5}{*}{Beauty} & HR@1 & 0.0251 & 0.0472 & 0.0831 & 0.0924 & 0.0812 & 0.0982 & 0.0753 & 0.0164 & 0.1218 & \underline{0.1265} & 0.1001 & \textbf{0.1359}$^\ast$ & 7.43\%\\ 
    ~ & HR@5 & 0.0858 & 0.1195 & 0.1569 & 0.2062 & 0.1780 & 0.1956 & 0.1845 & 0.0525 & 0.2329 & \underline{0.2359} & 0.2000 & \textbf{0.2511}$^\ast$ & 6.44\%\\
    ~ & HR@10 & 0.1474 & 0.1823 & 0.2112 & 0.2801 & 0.2489 & 0.2634 & 0.2633 & 0.1073 & 0.3000 & \underline{0.3027} & 0.2666 & \textbf{0.3225}$^\ast$ & 6.54\%\\
    ~ & NDCG@5 & 0.0553 & 0.0837 & 0.1213 & 0.1509 & 0.1309 & 0.1484 & 0.1311 & 0.0338 & 0.1796 & \underline{0.1831} & 0.1519 & \textbf{0.1957}$^\ast$ & 6.88\%\\
    ~ & NDCG@10 & 0.0751 & 0.1039 & 0.1387 & 0.1746 & 0.1536 & 0.1702 & 0.1565 & 0.0513 & 0.2012 & \underline{0.2046} & 0.1732 & \textbf{0.2186}$^\ast$ & 6.84\%\\
    \midrule
    \multirow{5}{*}{Sports} & HR@1  & 0.0186 & 0.0306 & 0.0525 & 0.0643 & 0.0281 & 0.0623 & 0.0561 & 0.0133 & 0.0811 & \underline{0.0865} & 0.0633 & \textbf{0.0954}$^\ast$ & 10.29\%\\ 
    ~ & HR@5 & 0.0750 & 0.0998 & 0.1263 & 0.1851 & 0.0901 & 0.1710 & 0.1876 & 0.0578 & 0.2051 & \underline{0.2061} & 0.1654 & \textbf{0.2208}$^\ast$ & 7.13\%\\
    ~ & HR@10 & 0.1385 & 0.1677 & 0.1921 & 0.2825 & 0.1527 & 0.2599 & \underline{0.2989} & 0.1072 & 0.2956 & 0.2964 & 0.2453 & \textbf{0.3161}$^\ast$ & 5.75\%\\
    ~ & NDCG@5 & 0.0464 & 0.0652 & 0.0894 & 0.1252 & 0.0589 & 0.1170 & 0.1219 & 0.0352 & 0.1441 & \underline{0.1473} & 0.1149 & \textbf{0.1593}$^\ast$ & 8.15\%\\
    ~ & NDCG@10 & 0.0667 & 0.0869 & 0.1105 & 0.1565 & 0.0789 & 0.1455 & 0.1577 & 0.0509 & 0.1731 & \underline{0.1764} & 0.1406 & \textbf{0.1899}$^\ast$ & 7.65\%\\
    \midrule
    \multirow{5}{*}{Movielens} & HR@1 & 0.0532 & 0.0965 & 0.0979 & 0.0653 & 0.1208 & 0.1273 & \underline{0.1274} & 0.0226 & 0.1207 & OOM & 0.0360 & \textbf{0.1345}$^\ast$ & 5.57\%\\
    ~ & HR@5 & 0.1954 & 0.2910 & 0.2992 & 0.2593 & 0.3362 & 0.3444 & \underline{0.3561} & 0.0847 & 0.3503 & OOM & 0.1456 & \textbf{0.3724}$^\ast$ & 4.58\%\\
    ~ & HR@10 & 0.3101 & 0.4266 & 0.4365 & 0.4286 & 0.4831 & 0.4854 & 0.4976 & 0.1504 & \underline{0.4979} & OOM & 0.2760 & \textbf{0.5230}$^\ast$ & 5.04\%\\
    ~ & NDCG@5 & 0.1244 & 0.1949 & 0.2002 & 0.1615 & 0.2304 & \underline{0.2385} & 0.2355 & 0.0534 & 0.2377 & OOM & 0.0898 & \textbf{0.2565}$^\ast$ & 7.54\%\\
    ~ & NDCG@10 & 0.1613 & 0.2386 & 0.2445 & 0.2159 & 0.2777 & 0.2840 & \underline{0.2877} & 0.0744 & 0.2853 & OOM & 0.1315 & \textbf{0.3051}$^\ast$ & 6.05\%\\
    \bottomrule
    \end{tabular}
    }
    \vspace{-0.1in}
\end{table*}

In this section, we carry out comprehensive experiments in various settings to address the following research questions:
\begin{itemize}[leftmargin=*]
\item \textbf{RQ1}: How does \model\ perform compare with state-of-the-arts?
\item \textbf{RQ2}: Can our adaptive contrastive learning paradigms improve the performance of sequential recommendation in various scenarios, such as cold-start users and sparse items?
\item \textbf{RQ3}: How do different parameters impact \model's performance?
\item \textbf{RQ4}: Can the effects of our debaising CL be explained?
\end{itemize}

\subsection{Experimental Setting}
\subsubsection{\bf Datasets} 
We evaluate our model using four public datasets sourced from three real-life platforms, \ie, Reddit, Amazon, and MovieLens. i) \textbf{Reddit}: This dataset captures user interactions with subscribed topics on the Reddit platform. ii) \textbf{Amazon}: This product dataset collects user-item interactions from Amazon with the categories of \textbf{Beauty} and \textbf{Sports} products. iii) \textbf{MovieLens-20M}: This dataset comprises rating behaviors gathered from a movie review website. The statistics for various datasets are provided in Table~\ref{tab:dataset}.


\vspace{-0.05in}
\subsubsection{\rm \textbf{Evaluation Protocols}}
We follow \cite{bert4rec, mbht, sasrec} to adopt the \textit{leave-one-out} strategy for model evaluation. Specifically, we treat the last interaction of each user as testing data, and designate the previous one as validation data. We employ the commonly utilized Hit Ratio (HR@N) and Normalized Discounted Cumulative Gain (NDCG@N) metrics, with N values of 1, 5, and 10.



\vspace{-0.05in}
\subsubsection{\rm \textbf{Baselines}} The compared methods are described as follows:

\paratitle{Non-GNN Sequential Recommendation Methods.}
\begin{itemize}[leftmargin=*]
    \item \textbf{Caser}~\cite{caser}. It employs CNN layers in both vertical and horizontal perspectives to capture the sequential information.
    \item \textbf{GRU4Rec}~\cite{gru4rec}. It employs GRU to encode sequences and incorporates a ranking-based loss for session-based recommendation.
    \item \textbf{SASRec}~\cite{sasrec}. This method is the pioneer in utilizing the self-attention to capture dynamic user interests within a sequence.
    \item \textbf{BERT4Rec}~\cite{bert4rec}. The Cloze task is introduced to sequential recommendation, employing a bidirectional attentive encoder.
\end{itemize}
\paratitle{Graph-based Sequential Recommender Systems.}
\begin{itemize}[leftmargin=*]
    \item \textbf{SR-GNN}~\cite{srgnn}. It produces hybrid embeddings that effectively represent both local and global user interests with graphs.
    \item \textbf{GCSAN}~\cite{gcsan}. It conducts self-attention on graph-based sequential embeddings to capture long-term user interests.
    \item \textbf{SURGE}~\cite{surge}. This approach incorporates metric learning to create a parameterized item similarity graph and leverages hierarchical attention to combine various aspects of user interests.
\end{itemize}
\paratitle{Self-Supervised Sequential Recommendation Models}.
\begin{itemize}[leftmargin=*]
    \item \textbf{S$^3$-Rec}~\cite{s3rec}. This approach develops self-supervised task over item sequences employing a pretrain-finetuning strategy.
    \item \textbf{CL4SRec}~\cite{cl4srec}. It empowers recommendation with different sequence-level augmentations \ie item crop, mask, and reorder.
    \item \textbf{DuoRec}~\cite{duorec}. This research investigates the representation degeneration issue in sequential recommendation and offers solutions based on contrastive learning techniques.
    \item \textbf{ICLRec}~\cite{iclrec}. This approach improves sequential recommendation by conducting clustering and contrastive learning on user intentions to enhance recommendation.
\end{itemize}

\vspace{-0.1in}
\subsubsection{\rm \textbf{Parameter Settings}} We implement our \model~ and most of the baselines with the \textit{RecBole}~\cite{recbole} library. For \model, the number of Transformer layer and GNN layer is set as 2. The embedding size is set as 64. $\mu_c$ controls the mean value of conformity scores is search from $[0.3, 0.4, 0.5, 0.6, 0.7]$ and $\sigma$ for the standard deviation is set as $0.1$. The weight $\lambda_1$ for the self-supervised learning loss is searched from $[$5e-4, 1e-3, 5e-3, 1e-2$]$ and $\lambda_2$ for the Kullback-Leibler divergence loss is searched from $[$1e-3, 1e-2, 1e-1, 1$]$.

\vspace{-0.1in}
\subsection{RQ1: Overall Performance}
We present the performance of our model and baselines in Table~\ref{tab:results}. Based on the results, we can make the following observations:
\begin{itemize}[leftmargin=*]
    \item Graph-based sequential recommendation models \ie SR-GNN, GCSAN, and SURGE achieve better overall performance compared with non-GNN models. The improvement is attributed to the effective capture of global item dependencies and long-term user interests facilitated by graph convolutions. Nonetheless, it is worth noting that the performance improvement is less significant or even negative in certain cases (e.g., SR-GNN on the Sports dataset) when dealing with sparser datasets such as Beauty and Sports. This limitation suggests that constructing graphs for sparse data may be inadequate for modeling long-term semantics, as a result of data scarcity and the presence of noise. \\\vspace{-0.12in}
    
    \item Sequential models incorporating self-supervised learning components, such as S$^3$-Rec, CL4SRec, DuoRec, and ICLRec, exhibit varying performance outcomes across the four datasets. For instance, CL4SRec and DuoRec demonstrate similar performance levels that surpass other baseline models on the Beauty and Sports datasets. On the other hand, S$^3$-Rec exhibits inferior results compared to other models across all four datasets. It is important to mention that this approach employs sequence augmentation and contrastive learning during the pretraining phase. In contrast to CL4SRec, which pursues the same objective during the main task training, this suggests that pretraining a sequence-level contrastive goal may not bring much benefits.\\\vspace{-0.12in}
    
    \item In comparison to the baseline models, our proposed method consistently outperforms them across the four datasets in general, with a particularly notable improvement in HR@1. While our method is marginally and not statistically significantly outperformed by SR-GNN in HR@5 and HR@10 on the Reddit dataset, the results in other cases still suggest the effectiveness of \model.
    
\end{itemize}

\subsection{RQ2: Benefits Study}
\subsubsection{\rm \textbf{Performance on Cold-Start Users}}

\begin{figure}[t]
\centering
\subfigure[HR@1]{
\label{fig:cold_user:hr}
\includegraphics[width=0.45\linewidth]{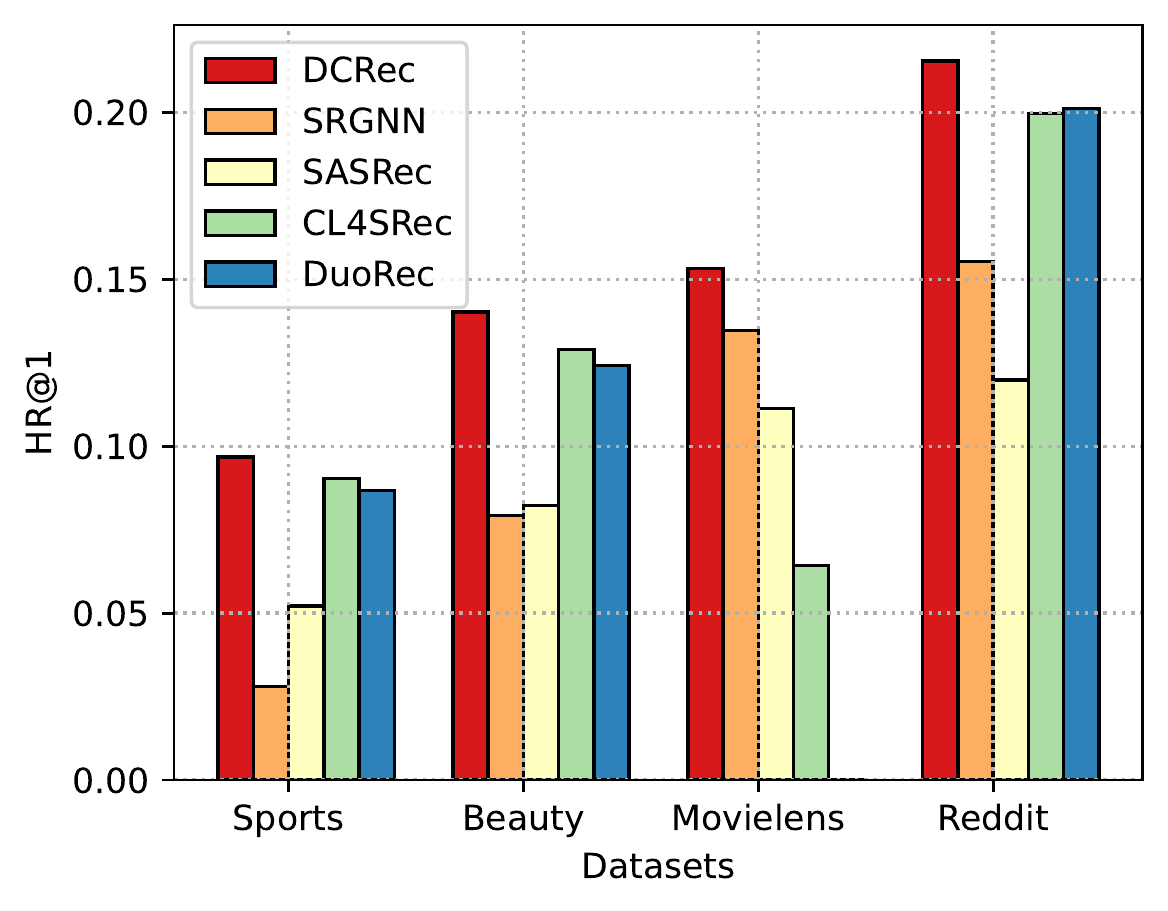}}
\subfigure[NDCG@5]{
\label{fig:cold_user:ndcg}
\includegraphics[width=0.45\linewidth]{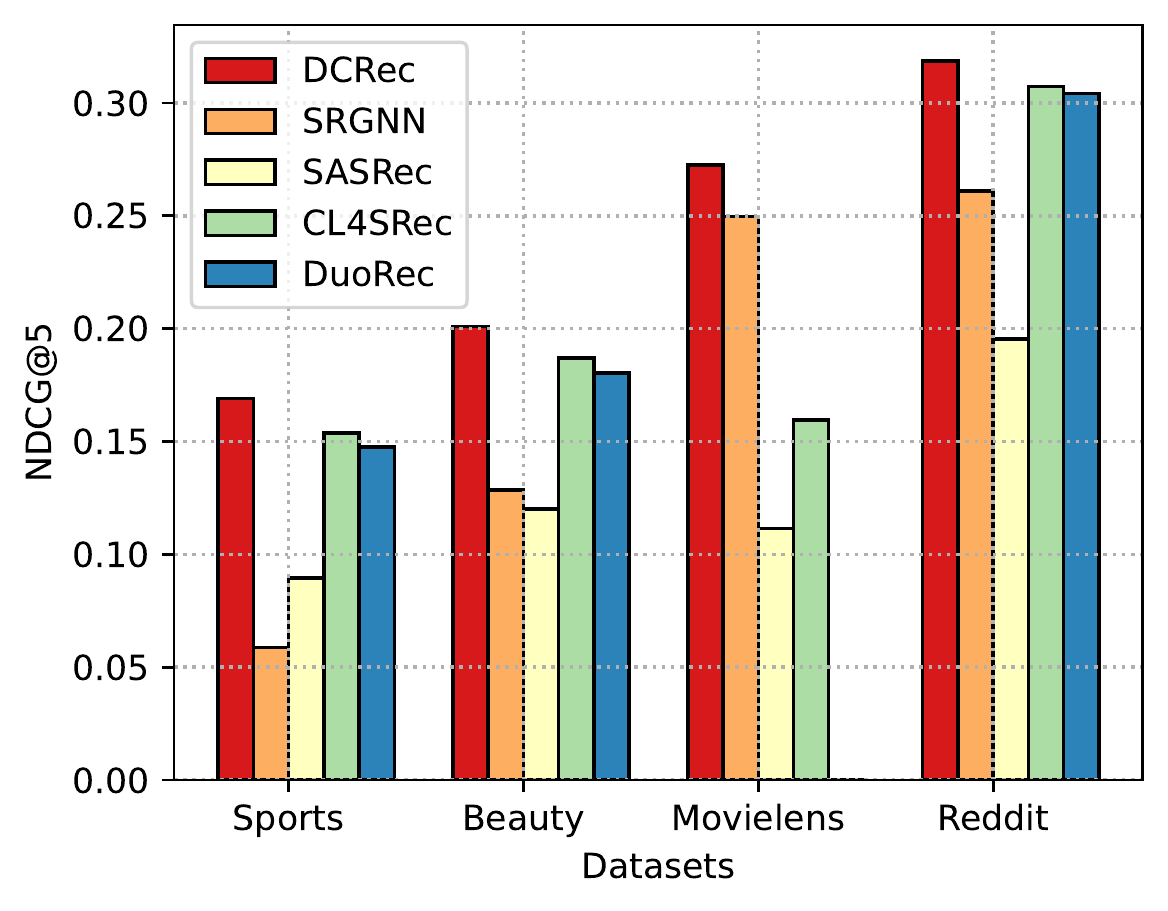}}
\vspace{-0.2in}
\caption{Evaluation results on cold-start users.}
\label{fig:cold_user}
\vspace{-0.25in}
\end{figure}

Adhering to evaluation settings outlined in~\cite{insert, mhcn, kgcl}, we filter users with fewer than 20 interactions to create a sub-dataset of cold-start users for the four datasets. The evaluation outcomes for cold-start users, including HR@1 and NDCG@5, are presented in Figure~\ref{fig:cold_user}. The results clearly demonstrate that our method outperforms the strongest baselines from various research lines in addressing the cold-start problem. We attribute this advantage to our model's capability to balance knowledge transfer across different views, utilizing conformity-aware contrastive learning. For cold-start users, the designed model effectively extracts valuable information from global transition signals and collaborative patterns to enhance user representations. Furthermore, it refines the acquired knowledge through conformity-aware weighting to mitigate the popularity bias affecting non-active users.


\subsubsection{\rm \textbf{Performance w.r.t. Item Sparsity}} 
\begin{figure}[t]
\centering
\subfigure[Sports Dataset]{
\label{fig:cold_item:hr}
\includegraphics[width=0.47\linewidth]{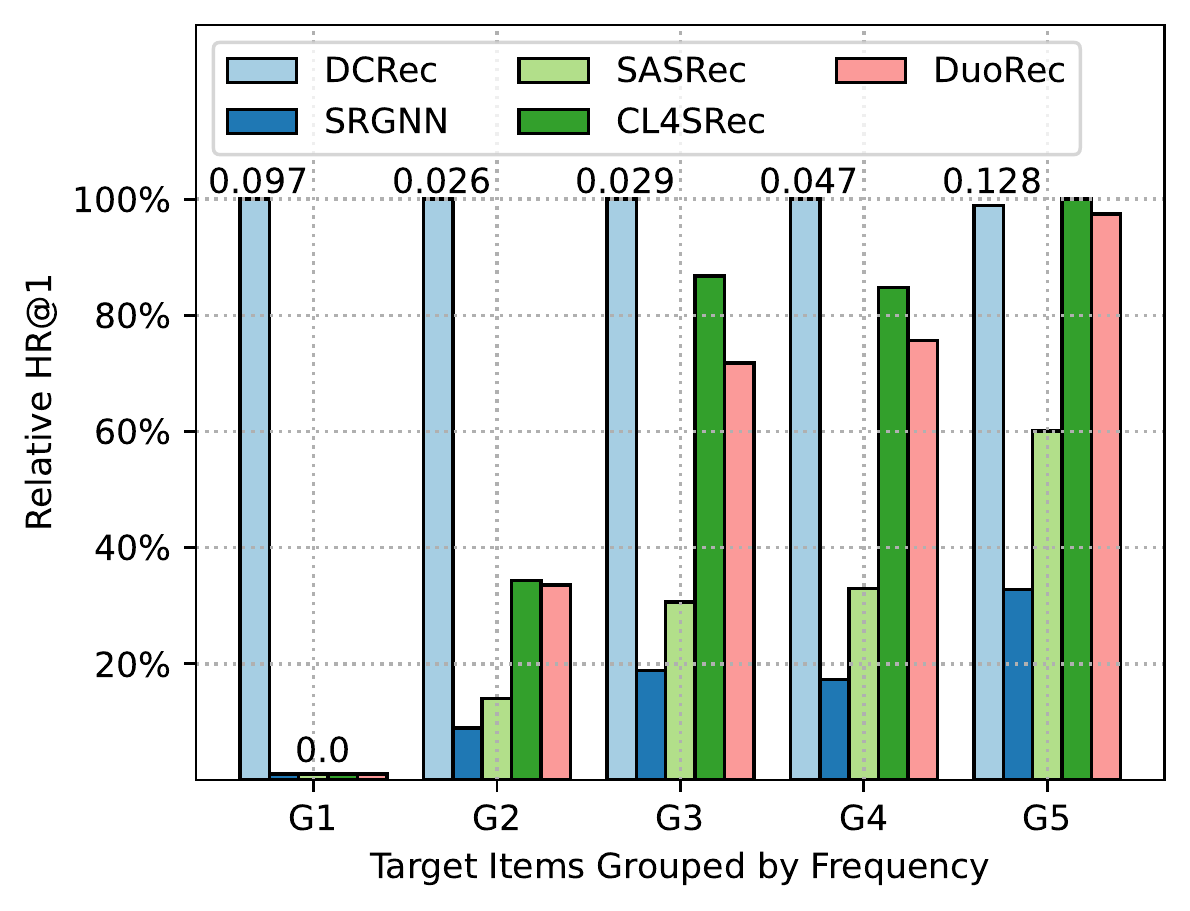}}
\subfigure[Beauty Dataset]{
\label{fig:cold_item:ndcg}
\includegraphics[width=0.47\linewidth]{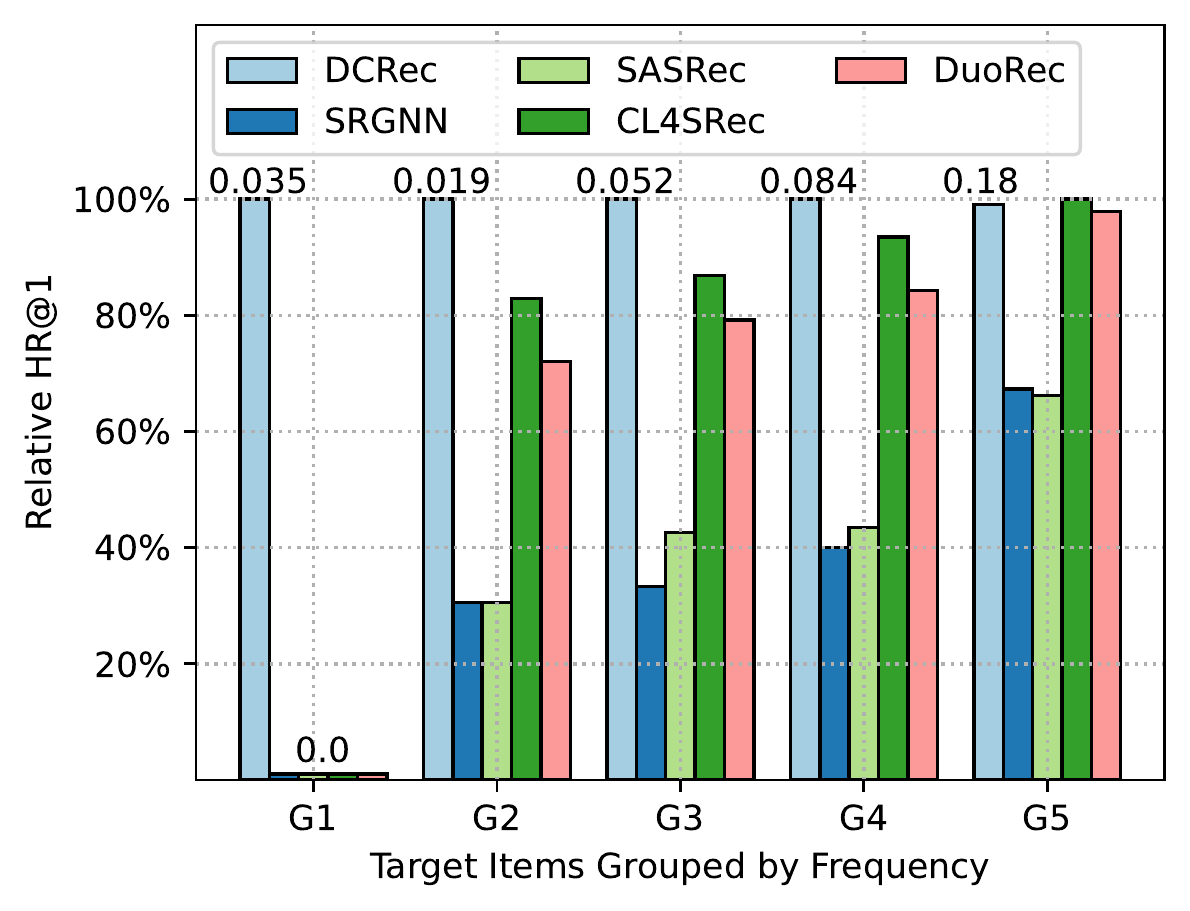}}
\vspace{-0.2in}
\caption{Performance on item groups w.r.t. sparsity level. Larger group number indicates more popular items.}
\label{fig:cold_item}
\vspace{-0.15in}
\end{figure}

To further explore our model's capabilities in addressing the item sparsity challenge in sequential recommendation, we categorize target items into five groups based on their sparsity levels. A lower group number indicates that the items within that group have fewer interactions with users. The results are displayed in Figure~\ref{fig:cold_item}. Across all five groups, our model outperforms the baselines in the first four groups, demonstrating its effectiveness in handling item sparsity. This observation suggests that the performance improvement of our model primarily stems from accurately predicting less popular items. Consequently, we posit that our debiased contrastive learning generates higher-quality item embeddings for recommendation. This conclusion aligns with our findings presented in Section \ref{sec:a:emb}.


\subsection{RQ3: Ablation Study}

\subsubsection{\rm \textbf{Impact of Key Components.}}
We develop four variants, with each one excluding a specific key component, to delve deeper into the design of our \model. Details are presented as follows:\vspace{-0.05in}
\begin{itemize}[leftmargin=*]
    \item \textbf{w/o T-CL} removes the contrastive learning between the sequential and item transition graph representations for augmentation.
    \item \textbf{w/o C-CL} removes the contrastive learning between the sequential and co-interaction graph representations for augmentation.
    \item \textbf{w/o CL} removes the entire contrastive learning module.
    \item \textbf{w/o Adaptive-CL} omits the design described in Section \ref{sec:adaptive} that enables the contrastive learning process to adapt based on the user's conformity and interest disentanglement. \vspace{-0.05in}
\end{itemize}
The results of the ablation study are presented in Table~\ref{tab:ablation}. Based on the ablation study, we can make the following observations: 1) Each of these key components contributes substantially to the enhancement of the model's recommendation performance; 2) The contrastive learning between the sequential and item transition graph representations leads to more significant improvements in the model's performance. 3) On the Beauty dataset, eliminating adaptive weights results in a lower performance than that achieved without contrastive learning. This observation suggests that the adaptive-CL component serves a crucial role in mitigating the bias introduced by potentially inaccurate contrastive learning.


\begin{table}[t]
    \centering
    \caption{Ablation study results of \model.}
    \vspace{-0.15in}
	\resizebox{\linewidth}{!}{
    \begin{tabular}{lcccccc}
    \toprule
    \multirow{2}{*}{Ablation Settings} & \multicolumn{2}{c}{Reddit} & \multicolumn{2}{c}{Beauty} & \multicolumn{2}{c}{Sports}\\
    \cmidrule(lr){2-3}\cmidrule(lr){4-5}\cmidrule(lr){6-7}
    ~ & HR@1 & HR@5 & HR@1 & HR@5 & HR@1 & HR@5\\
    \midrule
    \model & 0.188 & 0.390 & 0.136 & 0.251 & 0.095 & 0.221\\
    \cmidrule(lr){1-7}
    w/o T-CL & 0.178 & 0.374 & 0.122 & 0.224 & 0.083 & 0.195\\
    w/o C-CL & 0.182 & 0.387 & 0.132 & 0.245 & 0.083 & 0.202\\
    w/o CL & 0.169 & 0.369 & 0.120 & 0.221 & 0.084 & 0.196\\
    w/o Adaptive-CL & 0.174 & 0.366 & 0.121 & 0.225 & 0.082 & 0.192\\
    \bottomrule
    \end{tabular}
    }
    \label{tab:ablation}
    \vspace{-0.15in}
\end{table}

\vspace{-0.05in}
\subsubsection{\rm \textbf{Sensitivity to Hyperparameters.}}
Owing to space constraints in main file, we relocate the discussion to Section~\ref{sec:a:hp}.

\vspace{-0.1in}
\subsection{RQ4: Case Study}
\begin{figure}
    \centering
    \includegraphics[width=0.9\linewidth]{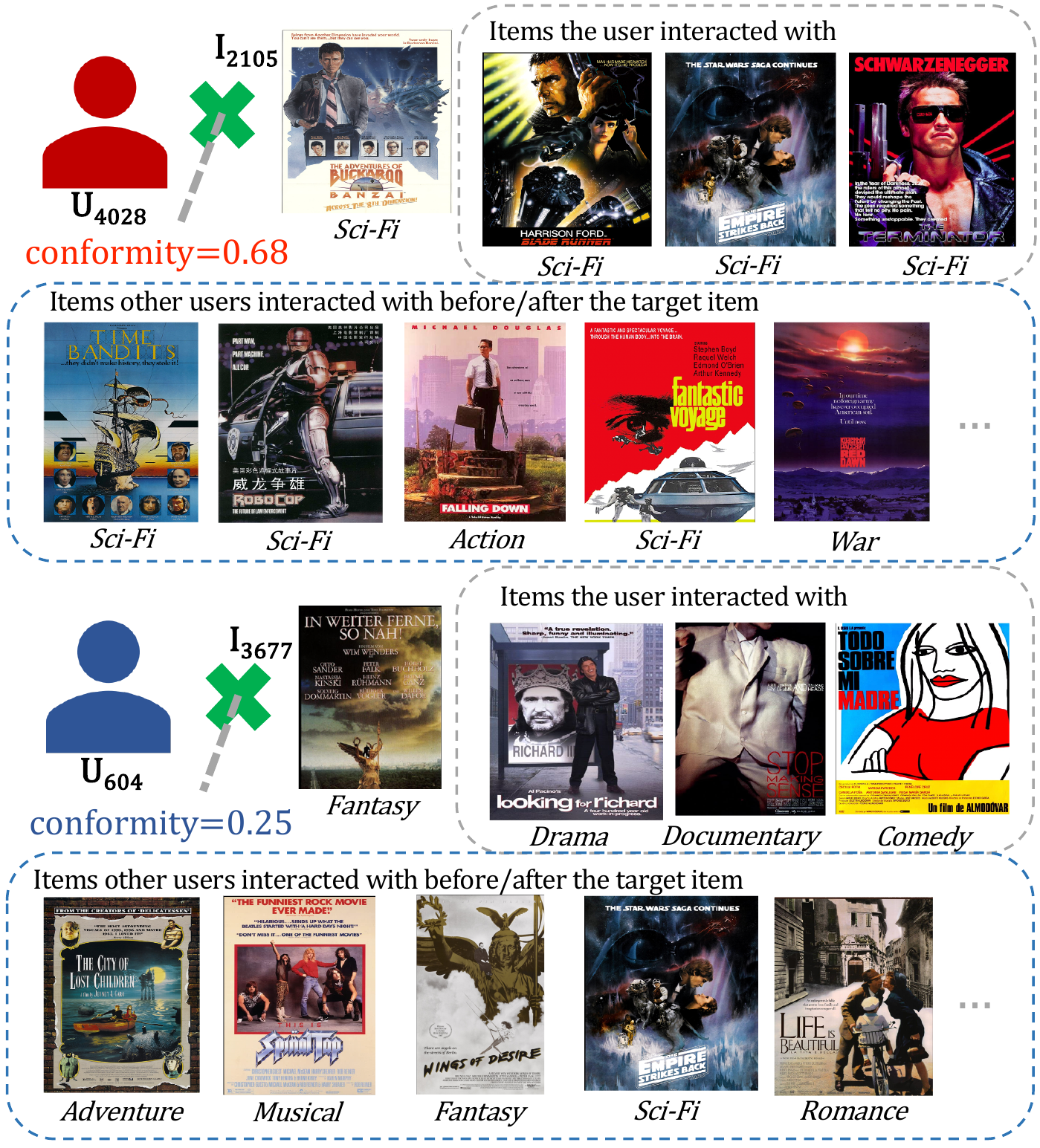}
    \vspace{-0.15in}
    \caption{Two user-item interaction pairs with different conformity levels discovered by \model\ from Movielens dataset.}
    \label{fig:case_study}
    \vspace{-0.15in}
\end{figure}

We conduct case studies on the Movielens to verify the rationality of conformity weights in our model. In Figure~\ref{fig:case_study}, we select two user-interaction pairs with different conformity degrees, specifically 0.68 and 0.25. In the first case, user $U_{4028}$ interacts with a sci-fi movie, and movies of the same genre are prevalent in their historical sequence. Additionally, we showcase movies from other users' sequences that are closely related to the target movie in order to visualize the transition graph. From the results, we see that most users interact with similar movie themes around the target movie, consistent with the target user's pattern. Hence, a conformity degree of 0.68 is a reasonable assessment of the user's conformity.


In the second sample, a user interacts with a fantasy movie, and her conformity degree is estimated as 0.25 by the model. This case suggests that the target user's preference has less in common with that of other users. Specifically, the user engaged with diverse movies across multiple genres. The interactions display no consistent semantic bias. As a result, we believe the interaction is guided by the user's authentic interest rather than the impact of popularity bias. Unlike the case in Figure~\ref{fig:intro_case}, we analyze disentanglement of user interest and conformity using item semantics, not popularity. Both factors are informative and jointly influence user intentions.



\section{Related Work}
\label{sec:relate}

\noindent \textbf{Sequential Recommendation}.
The advancement of neural networks and deep learning techniques has led to an increasing number of researchers proposing neural models to harness the rich latent semantics embedded in user behavior sequences. For instance, Caser~\cite{caser} relies on convolutional neural networks (CNNs), while GRU4Rec~\cite{gru4rec} is based on recurrent neural networks (RNNs). The subsequent introduction of Transformer~\cite{vaswani2017attention} has inspired researchers to develop sequential models like and BERT4Rec~\cite{bert4rec}, which use attention mechanisms to capture pairwise relations between user-interacted items. Moreover, recent GNN-based sequential models like SR-GNN~\cite{srgnn}, MTD~\cite{mtd}, and SURGE~\cite{surge} benefit from the strong capability of GNNs to capture global connections. \\\vspace{-0.12in}


\noindent \textbf{Self-Supervised Learning in Recommendation}.
Recently, self-supervised learning has become popular in recommender system research. In collaborative filtering (CF), SGL~\cite{sgl} uses random data augmentation on user-item graphs and applies self-discrimination contrastive learning on user/item nodes. SSL4Rec~\cite{ssl} employs data augmentation on item features and introduces a contrastive pre-training objective to improve learned representations in the two-tower model.
In knowledge-aware recommendation, KGCL~\cite{kgcl} develops a knowledge graph contrastive learning framework to aid denoising and integration between CF learning and knowledge graph encoding. For socially-aware recommendation, MHCN~\cite{mhcn} designs a graph infomax task to accommodate cascading semantic information from social graphs, enhancing user representation learning. In the field of sequential recommendation, CL4SRec~\cite{cl4srec} introduces sequential data augmentation into a contrastive learning task to derive more robust sequence representations. DuoRec~\cite{duorec} proposes a contrastive learning method based on sequence-level positive pairing to address the problem of representation degeneration in sequential recommenders. ICLRec~\cite{iclrec} conducts clustering and contrastive learning on user intents, and it enhances sequential recommendation by improving the representation of user interests.




\section{Conclusion}
\label{sec:conclusoin}

In this paper, our proposed new \model\ model discovers users' genuine interests from conformity to enhance sequential recommendation performance while mitigating popularity bias. The proposed debaised contrastive learning effectively reduces the impact of popularity bias in data augmentation for sequential recommender systems. Through comprehensive experiments, our \model\ has been shown to be effective, achieving superior results compared to other baselines. In our future work, it would be interesting to investigate methods for automatically searching neural parameters in conjunction with our interest and conformity disentanglement, with the aim of further improving our sequential recommender.


\section*{Acknowledgments}
This project is partially supported by 2022 Tencent Wechat Rhino-Bird Focused Research Program Research and Weixin Open Platform. This research work is also supported by Department of Computer Science \& Musketeers Foundation Institute of Data Science at the University of Hong Kong.

\clearpage
\bibliographystyle{ACM-Reference-Format}
\balance
\bibliography{sample-base}

\clearpage
\appendix \section{Appendix}
\label{sec:appendix}

\subsection{The Learning Process of \model}
The following section aims to provide additional details on the learning process of \model. In particular, we outline the algorithm steps in Algorithm \ref{algorithm}, which describe how \model\ propagates forward to compute loss in a batch training manner. Furthermore, we summarize the hyperparameters selected in our experiments that result in the different performance of \model. 



\begin{algorithm}
\SetKwInOut{Input}{Input}\SetKwInOut{Output}{Output}
\caption{The Learning Steps of \model}
\label{algorithm}
\Input{The item sequences of all users $\mathcal{S} = \{ \boldsymbol{s}_1, \boldsymbol{s}_2, \cdots, \boldsymbol{s}_{|U|} \}$, each temporal sequence defined as $\boldsymbol{s}_u=\left(v_1, v_2, \cdots, v_T\right)$.}
\Output{The overall training loss $\mathcal{L}$ to back propagate.}
\textbf{Build Graphs}\;
Build the adjacency matrix $\mathbf{A}_{\mathcal{G}_t} \in \mathbb{R}^{|\mathcal{V}| \times |\mathcal{V}|}$ for item transition graph $\mathcal{G}_t$ as in Equation~\ref{eq:gt}\;
Build the adjacency matrix for item co-interaction graph as $\mathbf{A}_{\mathcal{G}_c} = \mathbf{R}^\trans\mathbf{R}$\;
\textbf{Sample a batch of users $u \in \mathcal{B}_u$}\;
Perform graph convolutional function on $\mathcal{G}_t$ and $\mathcal{G}_c$ as in Equation 6 to generate item embeddings $\mathbf{X}$ and $\mathbf{Z}$ reflecting of transitional and co-interaction patterns respectively\;
Encode user sequence $\boldsymbol{s}_u$ to $\mathbf{H}_u$ from the sequential pattern following Equation 1-4\;
Mask user-interacted items to derive the augmented transition graph $\bar{\mathbf{A}}_{\mathcal{G}_c}$\;
Generate interaction-level conformity weights $\omega_{(u,v)}$ as in Equation~\ref{eq:omega}\;
Compute $\mathcal{L}_w$ for constraining the distribution of $\omega$ as in Equation~\ref{eq:kl}\;
\textbf{Conformity-aware Contrastive Learning}\;
Perform contrastive learning between $\mathbf{H}$ and $\mathbf{X}$, weighted by $\omega$ as in Equation~\ref{eq:cl1}. Returns loss $\mathcal{L}_u$\;
Perform contrastive learning between $\mathbf{X}$ and $\mathbf{Z}$, weighted by $\psi_{\left(u,v\right)}=1-\omega_{\left(u,v\right)}$ as in Equation~\ref{eq:cl2}. Returns loss $\mathcal{L}_v$\;
\textbf{View Aggregation and Training}\;
Fuse the three views $\mathbf{H},\mathbf{X},\mathbf{Z}$ to obtain the final item representations $\mathbf{p}_v$ as in Equation 11.
Calculate the recommendation loss $\mathcal{L}_{rec}$ following Equation 12\;
Unify the overall loss by multi-task training: $\mathcal{L} = \mathcal{L}_{rec} + \lambda_1\left(\mathcal{L}_u+\mathcal{L}_v\right) + \lambda_2\left(\mathcal{L}_w\right)$\;
\Return{$\mathcal{L}$}\;
\end{algorithm}


\subsection{Supplementary Experiments}
\subsubsection{\rm \textbf{Hyperparameter Sensitivity}}
\begin{figure*}[h]
\centering
\subfigure[Top co-interaction size $k$]{
\label{fig:hp:sasrec}
\includegraphics[width=0.29\linewidth]{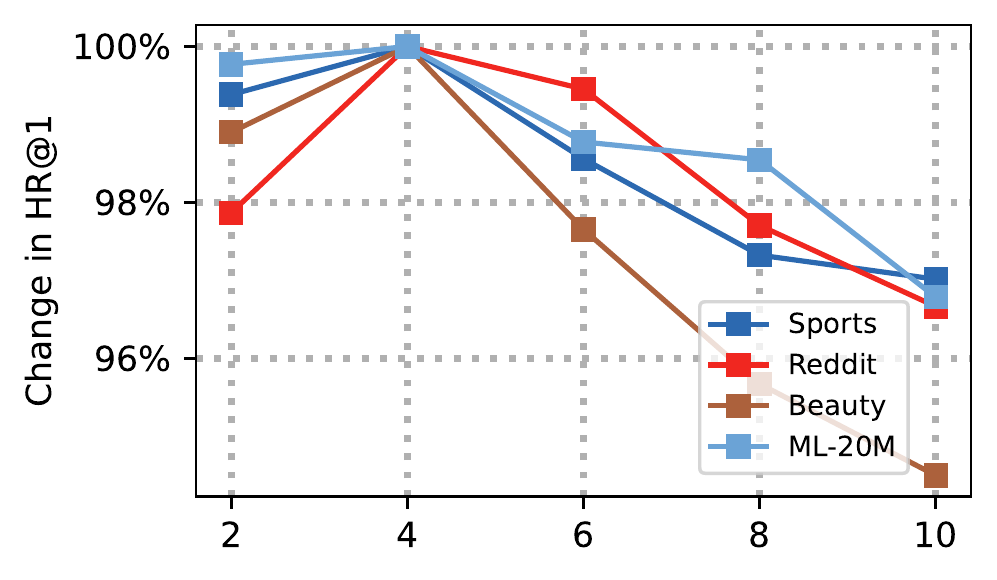}}
\subfigure[Mean of conformity $\mu_c$]{
\label{fig:hp:cl4srec}
\includegraphics[width=0.29\linewidth]{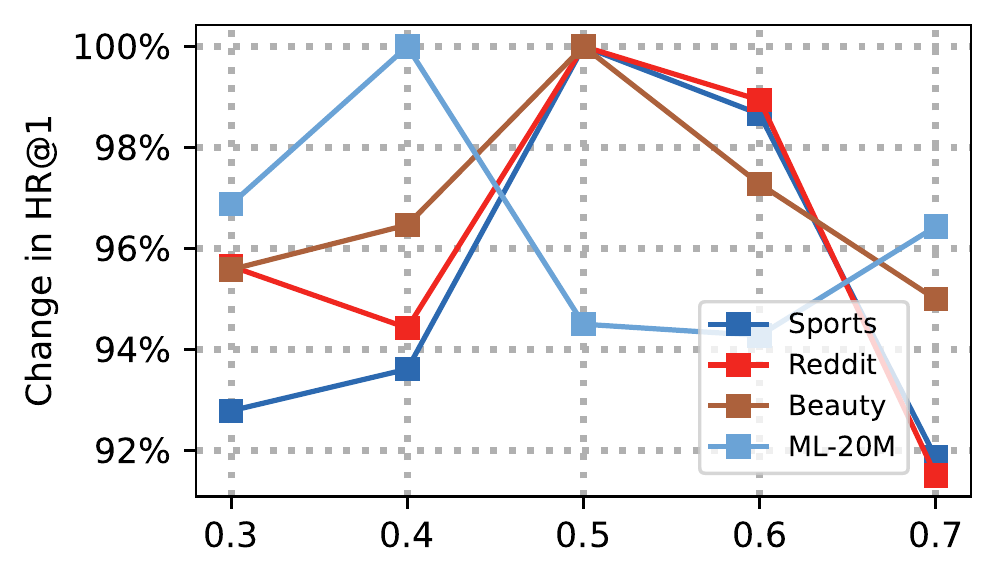}}
\subfigure[CL temperature $\tau$]{
\label{fig:hp:duorec}
\includegraphics[width=0.29\linewidth]{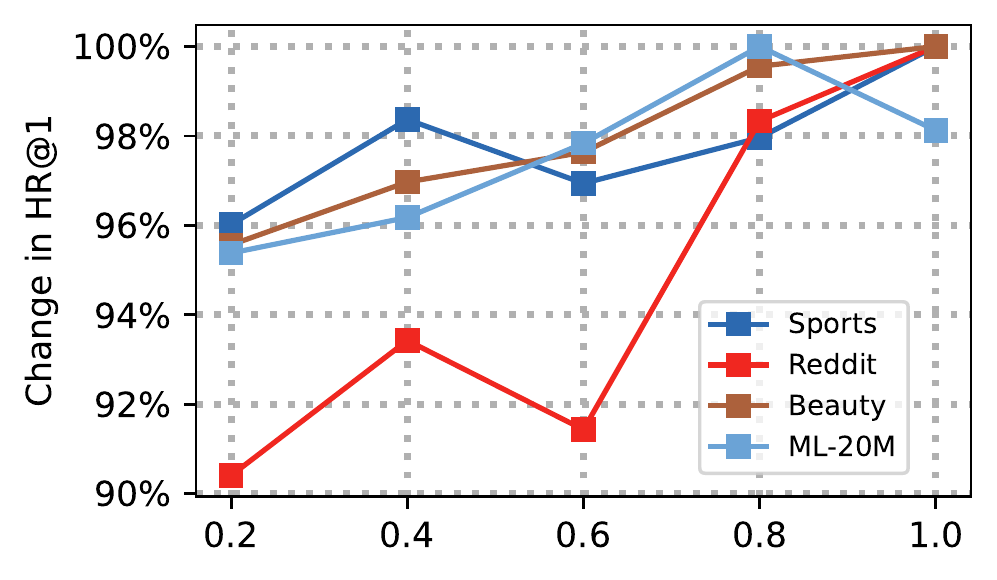}}
\vspace{-0.1in}
\caption{Hyperparameter study for \model~in terms of performance change with HR@1.}
\label{fig:hp}
\vspace{-0.1in}
\end{figure*}

We investigate the sensitivity of \model's performance with respect to different settings of key hyperparameters, including the top co-interaction size $k$ in $\mathcal{G}_c$, the mean of conformity weights $\mu_c$, and the temperature $\tau$ for contrastive learning. We conducted experiments on the four datasets by adjusting one hyperparameter within a specific range at a time, while keeping all others fixed. The evaluation results of our parameter study are presented in Figure~\ref{fig:hp}.


Based on the results, we summarize the following observations: (i) The performances generally first increase and then decrease as the top co-interaction size $k$ ranges from 2 to 10, with the best performance achieved all at 4. Increasing the top co-interaction size $k$ can bring more useful collaborative signals that boost performance. However, as $k$ continues to increase, it may introduce more noisy signals that are less relevant, causing the performance to decrease. (ii) We observed that the model's performance is sensitive to the $\mu_c$ hyperparameter, and that the best values are 0.4 or 0.5 across the four datasets. Since $\mu_c$ characterizes the average conformity degree across all users, we recommend adjusting it carefully for different datasets to match the specific user distribution. (iii) The results suggest that the best settings of $\tau$ are reported closer to 1.0. This observation is consistent with findings in previous work such as~\cite{duorec, iclrec}. A lower $\tau$ indicates a more differentiated contribution of common and hard samples~\cite{kgcl,sgl,khosla2020supervised}. We speculate that it is more difficult to obtain accurate hard negatives in sequential recommendation, which may explain why a higher $\tau$ reduces the negative impact of noisy samples for contrastive learning.


\label{sec:a:hp}

\subsubsection{\rm \textbf{Quality of Learned Item Embedding}}
\label{sec:a:emb}
\begin{figure}[]
\centering
\subfigure[SASRec]{
\label{fig:emb:sasrec}
\includegraphics[width=0.38\linewidth]{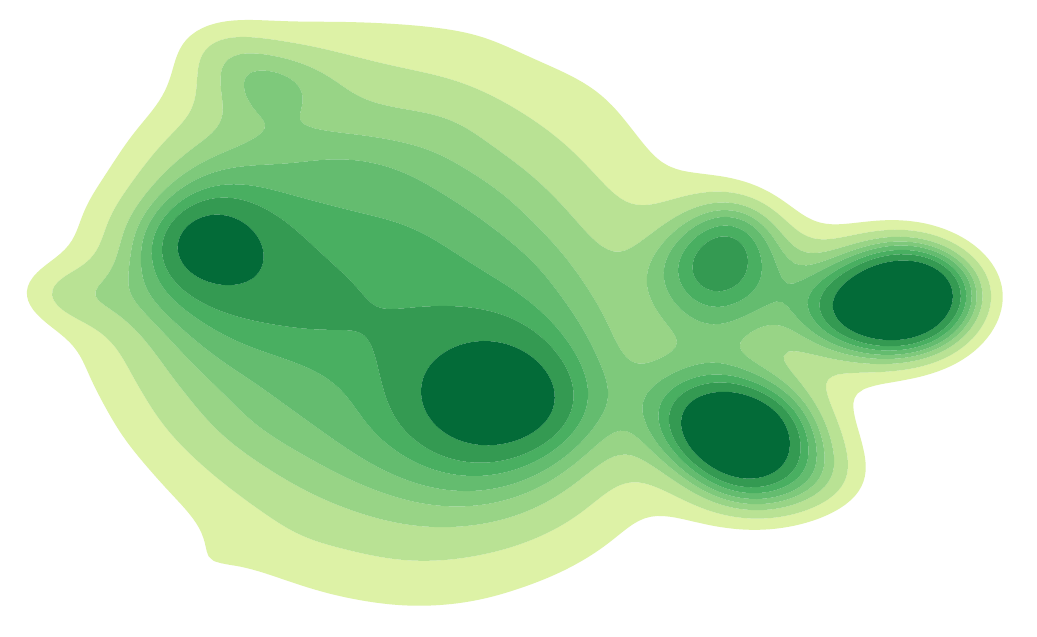}}
\subfigure[CL4SRec]{
\label{fig:emb:cl4srec}
\includegraphics[width=0.38\linewidth]{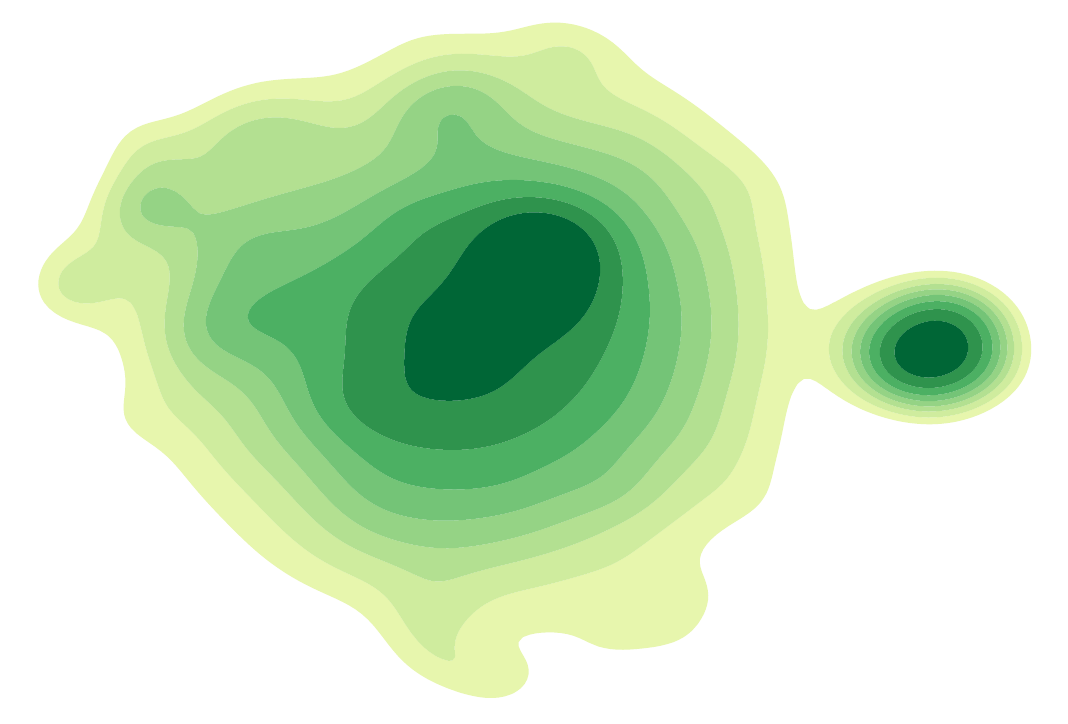}}
\subfigure[DuoRec]{
\label{fig:emb:duorec}
\includegraphics[width=0.38\linewidth]{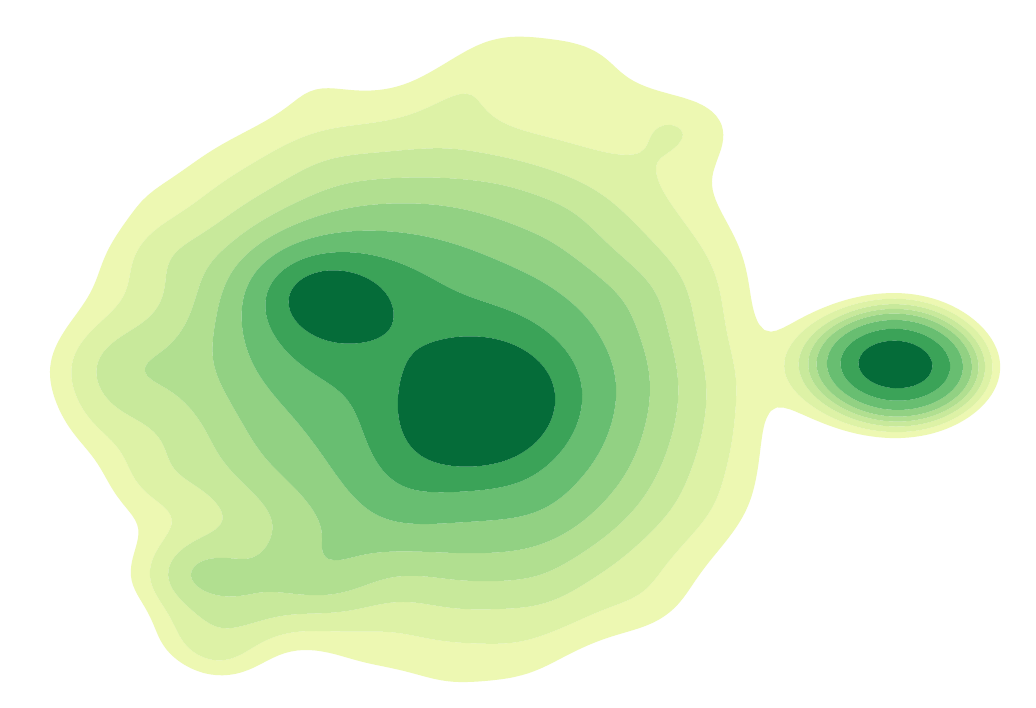}}
\subfigure[CLICD]{
\label{fig:emb:clicd}
\includegraphics[width=0.38\linewidth]{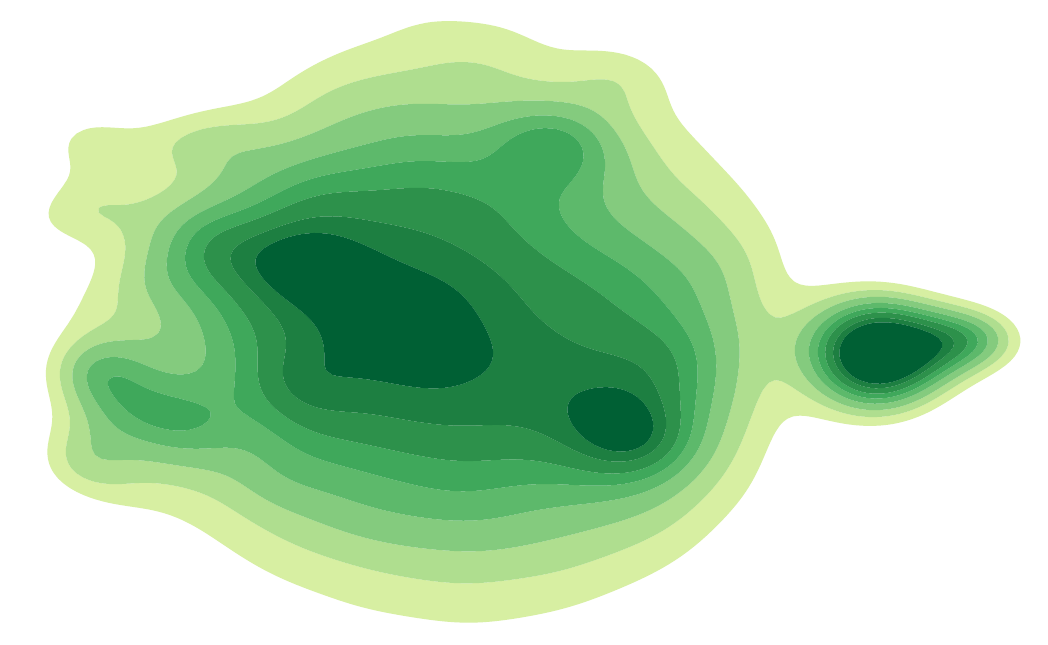}}
\vspace{-0.1in}
\caption{Item embedding visualization on Beauty dataset.}
\label{fig:emb}
\end{figure}

In this section, we demonstrate the superiority of the item embeddings learned by our proposed \model. We visualize the item embedding distribution learned by the several sequential baselines through 2-D KDE graphs. Using t-SNE and Gaussian kernel density estimation, we plot the embedding distribution of all items in the Beauty dataset, as shown in Figure~\ref{fig:emb}. The results reveal that, compared to other baselines, the item embeddings learned by \model~are more evenly and uniformly distributed. This even distribution provides better discrimination for user interests and item semantics.


\subsection{Formula Derivation Details}
In this section, we present the derivation of Equations \ref{eq:contrib1}-\ref{eq:contrib2}, which provide a measure of the contribution of positive and negative samples to the model learning. We begin by presenting the contrastive learning (CL) objective expressed in normalized vectors at the single interaction level:
\begin{equation}
    \mathcal{L}_u^{(u,v)} = - \omega_{\left(u,v\right)} \log \frac{\exp\left(\cos\left(\mathbf{\bar{h}}_v, \mathbf{\bar{x}}_v\right)/\tau\right)}{\sum_{v^\prime\in \mathcal{V}} \exp\left(\cos\left(\mathbf{\bar{h}}_v, \mathbf{\bar{x}}_{v^\prime}\right)/\tau\right)},
\end{equation}
\noindent where $\mathbf{\bar{h}}, \mathbf{\bar{x}}$ are normalized representations from two contrastive views. The gradient in terms of $\mathbf{h}_v$ can be expressed as:
\begin{equation}
    \nabla \mathcal{L}_u^{(u,v)} = \frac{\partial\mathcal{L}_u^{(u,v)}}{\partial\mathbf{\bar{h}}_v} \cdot \frac{\partial\mathbf{\bar{h}}_v}{\partial\mathbf{h}_v}.
\end{equation}
\noindent For the left term, we have:
\begin{equation}
    \begin{aligned}
    &\frac{\partial\mathcal{L}_u^{(u,v)}}{\partial\mathbf{\bar{h}}_v} = -\frac{\partial}{\partial\mathbf{\bar{h}}_v}\left(\mathbf{\bar{h}}_v^\trans\mathbf{\bar{x}}_v\right) + \frac{\partial}{\partial\mathbf{\bar{h}}_v}\log\sum_{v^\prime\in\mathcal{V}}\exp\left(\mathbf{\bar{h}}_v^\trans\mathbf{\bar{x}}_v{^\prime}\right)\\
    &= \frac{1}{\tau}\left(\frac{\sum_{v^\prime\in\mathcal{V}}\mathbf{\bar{x}}_{v^\prime}^\trans\exp\left(\mathbf{\bar{h}}_v^\trans\mathbf{\bar{x}}_v{^\prime}/\tau\right)}{\sum_{v^\prime\in\mathcal{V}}\exp\left(\mathbf{\bar{h}}_v^\trans\mathbf{\bar{x}}_v{^\prime}/\tau\right)}-\mathbf{\bar{x}}_v^\trans\right)\\
    &= \frac{1}{\tau}\left(\mathbf{\bar{x}}_v\frac{\exp\left(\mathbf{\bar{h}}_v^\trans\mathbf{\bar{x}}_v/\tau\right)}{\sum_{v^\prime\in\mathcal{V}}\exp\left(\mathbf{\bar{h}}_v^\trans\mathbf{\bar{x}}_{v^\prime}/\tau\right)} - \mathbf{\bar{x}}_v + \sum_{i\in V\setminus\{v\}} \mathbf{\bar{x}}_i^\trans \frac{\exp\left(\mathbf{\bar{h}}_v^\trans\mathbf{\bar{x}}_i/\tau\right)}{\sum_{v^\prime\in\mathcal{V}} \exp\left(\mathbf{\bar{h}}_v^\trans\mathbf{\bar{x}}_{v^\prime}/\tau\right) } \right)
    \end{aligned}
\end{equation}

Let
\begin{equation}
    P_{vi} = \frac{\exp\left(\mathbf{\bar{h}}_v^\trans\mathbf{\bar{x}}_i / \tau\right) }{ \sum_{i\in V\setminus\{v\}}\exp\left(\mathbf{\bar{h}}_v^\trans\mathbf{\bar{x}}_i / \tau \right) }
\end{equation}

We further derive Equation 19 as:
\begin{equation}
    \frac{\partial\mathcal{L}_u^{(u,v)}}{\partial\mathbf{\bar{h}}_v} = \frac{1}{\tau}\left(\mathbf{\bar{x}}_v^\trans\left(P_{vv}-1\right) + \sum_{i\in V\setminus\{v\}} \mathbf{\bar{h}}_i^\trans P_{vi} \right)
\end{equation}

For the right term, we have
\begin{equation}
\begin{aligned}
        \frac{\partial\mathbf{\bar{h}}_v}{\partial\mathbf{h}_v} &= \frac{\partial}{\partial\mathbf{h}_v}\left(\frac{\mathbf{h}_v}{\|\mathbf{h}_v\|}\right)\\
        &= \frac{1}{\|\mathbf{h}_v\|}\mathbf{I} + \mathbf{h}_v\left( \frac{\partial\frac{1}{\|\mathbf{h}_v\|}}{\partial\mathbf{h}_v} \right)\\
        &= \frac{1}{\|\mathbf{h}_v\|}\left(\mathbf{I}-\mathbf{\bar{h}}_v^\trans\mathbf{\bar{h}}_v\right)
\end{aligned}
\end{equation}
where $\mathbf{I}$ is the unit matrix.To this end, we have
\begin{equation}
\begin{aligned}
        \nabla \mathcal{L}_u^{(u,v)} &= \frac{1}{\tau\|\mathbf{\bar{h}}_v\|}\left(\mathbf{I}-\mathbf{\bar{h}}_v^\trans\mathbf{\bar{h}}_v\right)\left(\mathbf{\bar{x}}_v^\trans\left(P_{vv}-1\right) + \sum_{i\in V\setminus\{v\}} \mathbf{\bar{h}}_i^\trans P_{vi} \right)\\
        &=\frac{1}{\tau \|\mathbf{h}_v\|}\left(c(v)+\sum_{v^\prime\in V\setminus\{p\}}c(v^\prime)\right)
\end{aligned}
\end{equation}
which corresponds to Equation~\ref{eq:contrib1}-\ref{eq:contrib2}.

\end{document}